\documentclass[journal,onecolumn,12pt]{IEEEtran}

\ifCLASSINFOpdf
\else
\fi

\usepackage[utf8]{inputenc}
\usepackage{amssymb}
\usepackage[font=small]{caption}
\usepackage{amsmath}
\usepackage{comment}
\usepackage{mathtools}
\DeclarePairedDelimiter{\norm}{\lVert}{\rVert}
\usepackage{cite}
\usepackage{subfig}
\usepackage{graphicx}
\usepackage{multicol}
\usepackage{algorithm}
\usepackage[noend]{algpseudocode}
\usepackage{tikz}
\usepackage[english]{babel}
\usepackage{amsthm}
\theoremstyle{plain}
\usepackage{dirtytalk}
\newtheorem{definition}{Definition}
\usepackage{amsmath}
\DeclareMathOperator*{\argmax}{arg\,max}

\usepackage{mathtools}

\usepackage{tabularx}
\newcolumntype{L}[1]{>{\raggedright\arraybackslash}p{#1}}
\newcolumntype{C}[1]{>{\centering\arraybackslash}p{#1}}
\newcolumntype{R}[1]{>{\raggedleft\arraybackslash}p{#1}}
\usepackage{multirow}

\begin{document}
\title{Rate-Splitting Multiple Access for Downlink Multiuser MIMO: Precoder Optimization and PHY-Layer Design
\vspace{1.2cm}}
\author{\IEEEauthorblockN{Anup~Mishra, Yijie~Mao, \IEEEmembership{Member, IEEE}, Onur~Dizdar, \IEEEmembership{Member, IEEE}, and Bruno~Clerckx, \IEEEmembership{Senior Member, IEEE}}
\thanks{This work has been partially supported by the U.K. Engineering and Physical Sciences Research Council (EPSRC) under grant EP/N015312/1, EP/R511547/1.}}

\maketitle

\begin{abstract}
Rate-Splitting Multiple Access (RSMA) has recently appeared as a powerful and robust multiple access and interference management strategy for downlink Multi-user (MU) multi-antenna communications. In this work, we study the precoder design problem for RSMA scheme in downlink MU systems with both perfect and imperfect Channel State Information at the Transmitter (CSIT) and assess the role and benefits of transmitting multiple common streams. Unlike existing works which have considered single-antenna receivers (Multiple-Input Single-Output--MISO), we propose and extend the RSMA framework for multi-antenna receivers (Multiple-Input Multiple-Output--MIMO) and formulate the precoder optimization problem with the aim of maximizing the Weighted Ergodic Sum-Rate (WESR). Precoder optimization is solved using Sample Average Approximation (SAA) together with the proposed vectorization and Weighted Minimum Mean Square Error (WMMSE) based approach. Achievable sum-Degree of Freedom (DoF) of RSMA is derived for the proposed framework as an increasing function of the number of transmitted common and private streams, which is further validated by the Ergodic Sum Rate (ESR) performance using Monte Carlo simulations. Conventional MU-MIMO based on linear precoders and Non-Orthogonal Multiple Access (NOMA) schemes are considered as baselines. Numerical results show that with imperfect CSIT, the sum-DoF and ESR performance of RSMA is superior than that of the two baselines, and is increasing with the number of transmitted common streams. Moreover, by better managing the interference, RSMA not only has significant ESR gains over baseline schemes but is more robust to CSIT inaccuracies, network loads and user deployments.     
\end{abstract}

\IEEEpeerreviewmaketitle

\section{Introduction}\label{Intro}

\IEEEPARstart 
Multiple-Input Multiple-Output (MIMO) communication networks are one of the key enabling technologies for current and future wireless networks. By multiplexing signals in space, MIMO networks are capable of providing remarkably higher spectral efficiency. For a point-to-point MIMO channel, the channel capacity is known to scale linearly with the minimum number of transmit and receive antennas at high Signal-to-Noise Ratio (SNR), regardless of the level of Channel State Information (CSI) available to the Base Station (BS) \cite{DPCrateRegion03Goldsmith}. However, such result does not hold in a Multi-user (MU) network environment. When Channel State Information at the Transmitter (CSIT) is perfect, it is well known that Dirty Paper Coding (DPC) achieves the capacity region of the MIMO Broadcast Channel (BC) \cite{ZFandZFDPC2003,capacityRegion2006HW} but its implementation is prohibitive due to high computational complexity. A more practical approach that has attracted great attention is Space Division Multiple Access (SDMA) and MU--MIMO implemented using MU Linear Precoders (LP)\footnote{MU--MIMO techniques based on LP will be referred to as MU--MIMO in the rest of the paper.}\cite{wmmse08,MurchBD@2003,Sadek@Sayed2007,Zpan@Wong2004}. However, this approach has many limitations. Among them, one crucial limitation is that SDMA and MU-MIMO require accurate CSIT to design beamforming and manage interference. Therefore, it is
sensitive to the CSIT inaccuracy. In practical wireless communication networks, the CSIT quality can deteriorate due to several reasons such as pilot reuse in Time Division Duplex (TDD), mobility, latency, quantization errors in Frequency Division Duplex (FDD) and hardware impairments. Poor CSIT quality leads to poor interference management in current MIMO BC based on SDMA and MU--MIMO, and therefore, acts as a primary bottleneck in meeting demands of higher data rates.
\par Besides SDMA, another multiple access scheme that has recently attracted great attention in multi-antenna MU networks to manage interference is Non-Orthogonal Multiple Access (NOMA). Motivated by the established result in Single-Input Single-Output (SISO) BC, where power-domain NOMA based on Superposition Coding (SC) at the transmitter and Successive Interference Cancellation (SIC) at the receivers is known as a capacity-achieving scheme (such scheme is also known as SC--SIC), NOMA has been applied to multi-antenna BC \cite{QSunNOMA2015,QZhangNOMA2016}. A typical MIMO NOMA strategy 
is to group the users into different user groups and apply SIC within each group to decode the intra-group interference. The inter-group interference is treated as noise \cite{mao2017rate}. However, it has been pointed out in \cite{mao2017rate,hamdi2017bruno,bruno2019wcl,clerckx2021noma} that NOMA, which forces one user to decode the message of other users, causes spatial Degree of Freedom (DoF) loss and is inefficient in multi-antenna settings. Hence, in multi-antenna (Gaussian) BC with perfect CSIT, the only known capacity-achieving scheme is Dirty Paper Coding (DPC), rather than NOMA.
\par To overcome the limitations of SDMA and NOMA, a novel multiple access scheme, called Rate-Splitting Multiple Access (RSMA) is introduced in \cite{mao2017rate} for Multiple-Input Single-Output (MISO) BC. RSMA is based on splitting user messages into common and private parts at the transmitter. The common parts of the user messages are combined into common messages and encoded into common streams to be decoded by multiple users (but not necessarily intended to all those users). The private parts of all users are independently encoded into private streams to be decoded by the corresponding users only and treated as noise by other users. The encoded common and private streams are superposed in a non-orthogonal manner. Users rely on SIC to first decode the intended common streams before decoding the intended private streams. By adjusting the message split and the power allocation to the common and private streams, RSMA manages to partially decode the interference and treat the remaining interference as noise. This capability allows RSMA to act as a bridge between the two extreme interference management schemes of fully treating interference as noise (as in SDMA and MU–MIMO) and fully decoding interference (as in NOMA), and creates the opportunity to enhance the Quality of Service (QoS) and reduce complexity \cite{mao2019TCOM}. RSMA is shown to be a more general multiple access scheme embracing SDMA and NOMA as special cases \cite{mao2017rate,bruno2019wcl}. It is built upon 1-layer Rate-Splitting (RS), a low-complexity strategy that relies on one layer of SIC at each user \cite{RSintro16bruno}. For simplicity, 1-layer RS will be referred to by ‘RS’ for the rest of the paper.
\par The concept of RS was first introduced in \cite{TeHan1981} for a two-user SISO Interference Channel (IC) and has nowadays been further developed for multi-antenna BC \cite{bruno2019wcl,RS2015bruno,hamdi2016robust,enrico2016bruno,RS2016hamdi,chenxi2017bruno,chenxi2017brunotopology,Medra2018SPAWC,Lu2018MMSERS}. In multi-antenna BC, RS has been studied from an information theoretic perspective \cite{RSintro16bruno, RS2016hamdi,enrico2017bruno,chenxi2017bruno} demonstrating that it achieves the optimal sum-DoF \cite{RS2016hamdi} and the entire DoF region \cite{enrico2017bruno} of the $K$-user underloaded MISO BC with imperfect CSIT. \cite{chenxi2017bruno} investigates the achievable DoF region of RS in asymmetric MIMO BC and IC with imperfect CSIT and this achievable DoF region is shown in \cite{jafar@davoodiMIMODoF} to be optimal. Expanding the scope from high SNR to finite SNR regime of MISO BC, the energy efficiency performance of RSMA is investigated in \cite{mao2018EE} and its spectral efficiency performance is investigated in \cite{mao2017rate} with perfect CSIT and in \cite{RS2016hamdi,hamdi2016robust} with imperfect CSIT. RSMA has been shown to achieve a rate region close to DPC in MISO BC with perfect CSIT \cite{mao2017rate}. When CSIT is imperfect, linearly precoded RS is able to achieve a larger rate region than DPC \cite{mao2019beyondDPC}. A novel non-linearly precoded RS scheme, namely, Dirty Paper Coded RS is also proposed in \cite{mao2019beyondDPC}, which is shown to outperform linearly precoded RS and DPC. \cite{Flores2018ISWCS} has explored the benefits of RS using non-linear precoding technique named Tomlinson-Harashima Precoding (THP) in MISO BC. \cite{zheng@shamai} investigates precoder design and stream selection for RS in MISO BC. Apart from the conventional MISO BC, performance benefits of RS have also been exploited in massive MIMO \cite{Minbo2016MassiveMIMO,AP2017bruno}, millimeter wave systems \cite{minbo2017mmWave,Kola2018SPAWC}, multigroup multicasting \cite{hamdi2015multicasting,Tervo2018SPAWC}, multicarrier multigroup multicast \cite {chen@Xiao}, joint unicast and multicast transmission \cite{mao2019TCOM}, Cloud Radio Access Network (C-RAN) \cite{Ahmad2018SPAWC}, cooperative user relaying \cite{mao2019maxmin}, secure transmission \cite{hao@kwan}, etc. Furthermore, \cite{Lihua@Xingwang} investigates resource allocation for multicarrier RSMA systems, \cite{jaafer@naser} studies RSMA in aerial networks and RSMA is shown to provide better robustness, rate and QoS in multi-cell Coordinated Multipoint (CoMP) \cite{mao2018networkmimo}. RSMA is therefore a more promising strategy to manage interference in MIMO networks with both perfect and imperfect CSIT. 
\par RS has been fairly studied and analyzed in different aforementioned works. However, the number of receive antennas at each user is limited to one in most of these works. While there are studies considering multi-antenna receivers (in MIMO settings), the scope of such studies is limited. For example, \cite{8968439} proposes practical stream combining techniques together with Regularized Block Diagonalization (RBD) precoding for RS in MIMO BC but with only a single common stream and without precoding optimization. To the best of our knowledge, the role and benefits of multiple common streams in MIMO BC at finite SNR have not been investigated and remains an open problem. Moreover, in both perfect and imperfect CSIT settings, the achievable rate region of RS in MIMO BC is still unknown.

\subsection{Motivations and Contributions}
In light of the information theoretic results of \cite{chenxi2017bruno}, in a symmetric setup with $M$ transmit antennas and $Q$ receive antennas at each user, $\min(M,Q)$ common streams should be transmitted in RS to achieve the information theoretical optimal DoF with imperfect CSIT. Motivated by this result and the performance gain of RS over SDMA, NOMA and DPC in terms of rate and sum-DoF with imperfect CSIT in MISO BC, we fill the aforementioned research gaps and make the following contributions:
\begin{itemize}
\item We introduce a general framework of RS in symmetric MIMO BC with the same number of receive antennas at all users. The setting is general in the sense that RS can have arbitrary number of common streams between $1$ and $\min(M,Q)$ inclusive. This is the first work that allows flexibility in the number of common and private streams to be transmitted in RS.
\item At high SNR, we derive the achievable sum-DoF for the proposed RS framework in MIMO BC with imperfect CSIT and show the influence of multiple common streams on the sum-DoF of RS. Even with a single common stream, the sum-DoF of RS is shown to be greater than the sum-DoF of MU--MIMO and MIMO NOMA. This sum-DoF of RS increases as the number of common streams increases. We show that by transmitting multiple common streams, the sum-DoF gain of RS over MU--MIMO and MIMO NOMA increases. The assertions are further testified through the Ergodic Sum Rate (ESR) performance using Monte-Carlo simulations. This is the first paper to compare the DoF of MU-MIMO, MIMO NOMA and RS in a MIMO setting as opposed to current comparisons as in \cite{clerckx2021noma} which are limited to MISO settings.
\item We propose to utilize vectorization and Weighted Minimum Mean Square Error (WMMSE)-based Alternative Optimization (AO) algorithm to optimize the precoders for RS in MIMO BC with the aim of maximizing the WSR subject to the transmit power constraint. The proposed optimization framework addresses the challenge of intractable optimization introduced due to matrix variables. To the best of our knowledge, this is the first work that studies the precoder optimization and the benefits of transmitting multiple commons streams in RS-assisted MIMO BC with perfect and imperfect CSIT.
\item Under the assumption of Gaussian signalling and infinite block lengths, we demonstrate that the Ergodic Rate (ER) region of RS with optimized precoders always outperforms the ER regions of MU--MIMO and MIMO NOMA in MIMO BC with both perfect and imperfect CSIT. When CSIT is perfect, we also demonstrate that the ER region of RS comes closer to the capacity region achieved by DPC than MU--MIMO and MIMO NOMA. This is the first work to demonstrate such benefits of RS in MIMO settings.
\item To demonstrate the performance of RS in practical systems, we design the Physical (PHY)-layer architecture of RS with finite constellation modulation schemes, finite length polar codes and Adaptive Modulation and Coding (AMC). We show via the Link Level Simulations (LLS) that RS achieves significant throughput gain over MU--MIMO and MIMO NOMA in MIMO BC. This is the first work to design the PHY-layer architecture and to provide the LLS of RS in MIMO settings.
\end{itemize}
\subsection{Organisation}
The rest of the paper is organized as follows. In Section \ref{Sysmod}, the system model and CSIT assumptions are delineated. Problem is formulated in Section \ref{ProbForm}. Section \ref{OptFrame} contains the proposed methodology to solve the optimization problem. Section~\ref{sec:transceiver} describes the PHY-layer architecture for RS. Simulation results are illustrated in Section \ref{NumRes} and Section \ref{Concl} concludes the paper. Appendix A contains the derivation of the achievable sum-DoF for RS, MU-MIMO and MIMO NOMA schemes.
\subsection{Notations}
Matrices are denoted by boldface uppercase letters, column vectors are denoted by boldface lowercase letters and scalars are denoted by standard letters. Trace and determinant of matrix $\mathbf{A}$ are denoted by  $tr(\mathbf{A})$ and $\det(\mathbf{A})$, respectively. $diag(\mathbf{A})$ denotes the diagonal entries of the matrix. $\mathbf{A}^T$ and $\mathbf{A}^H$ denote the Transpose and Hermitian operators on the matrix $\mathbf{A}$, respectively. Euclidean norm of a vector $\mathbf{a}$ is denoted as $\norm{\mathbf{a}}$. $\otimes$ denotes the kronecker product and $vec(\mathbf{A})$ denotes vectorization of matrix $\mathbf{A}$. $\mathbb{E}_{X}\{Y\}$ is expectation of $Y$ w.r.t random variable $X$. $\mathbb{C}^{M\times N}$ and $\mathbb{R}^{M\times N}$ denote the set of all $M\times N$ dimensional matrices with complex-valued and real-valued entries, respectively. The Circularly Symmetric Complex Gaussian (CSCG) distribution with mean $\mu$ and variance $\sigma^{2}$ is denoted as $\mathcal{CN}(\mu,\sigma^{2})$.

\section{System Model}\label{Sysmod}
We consider a system model in which a BS consisting  of $M$ transmit antennas is serving $K$ users indexed by the set $\mathcal{K}= \{1,\ldots,K\}$, each equipped with $Q$ receive antennas. The transmit signal $\mathbf{x} \in \mathbb{C}^{M\times 1}$ is subject to a power constraint $\mathbb{E}\{\|\mathbf{x}\|^{2}\} \leq {P_{t}}$. The signal is transmitted through a MIMO BC with $\mathbf{H}_{k} \in \mathbb{C}^{M\times Q}$ denoting the channel matrix between the BS and user-$k$ and it is drawn from a continuous distribution. The signal received at user-$k$ is given by 
\begin{equation} \label{eq:sys1}
\mathbf{y}_{k} = \mathbf{H}_{k}^H\mathbf{x}+\mathbf{n}_{k},
\end{equation}
where $\mathbf{n}_k \sim \mathcal{C}\mathcal{N}({\mathbf{0}},\sigma_{n,k}^{2}\mathbf{I}_Q)$ is the Additive White Gaussian Noise (AWGN) vector and is independent of the channel matrices. Without loss of generality, we assume the noise variances across users to be equal, i.e., $\sigma_{n,k}^{2}=\sigma_{n}^{2},\,\forall k\in\mathcal{K}$. We assume that each user has complete knowledge of the channel information, i.e., perfect CSI at the Receiver (CSIR). In contrast, the BS only has partial knowledge of users CSI. Next we detail the channel acquisition at the BS.
\subsection{Imperfect CSIT}
\par The overall channel state can be denoted as
$\mathbf{H}=[\mathbf{H}_{1},\mathbf{H}_{2},\ldots,\mathbf{H}_{K}]$ $\in \mathbb{C}^{M\times (QK)}$, where the fading channel varies according to an ergodic stationary process during the time of transmission. The probability density function of the stationary process is ${f}_{\mathrm{\mathbf{H}}}(\mathbf{H})$. Practical limitations in CSI acquisition such as quantized feedback \cite{NJindalMIMO2006}, feedback and processing delay \cite{doppler2010Caire}, \cite{DoF2013SYang}, hardware impairments \cite{7892949} and channel estimation \cite{1312613} result in partial knowledge of the CSI at the BS given by $\widehat{\mathbf{H}}=[\widehat{\mathbf{H}}_{1},\widehat{\mathbf{H}}_{2},\ldots,\widehat{\mathbf{H}}_{K}]$ and is modeled as $\mathbf{H} = \widehat{\mathbf{H}} + \widetilde{\mathbf{H}}$.
We assume that the joint distribution of the channel state and its estimate $\{\mathbf{H},\widehat{\mathbf{H}}\}$ is ergodic and stationary \cite{RS2016hamdi}. The conditional density ${f}_{\mathrm{\mathbf{H}}|\mathrm{\widehat{\mathbf{H}}}}(\mathbf{H}|\widehat{\mathbf{H}})$ is assumed to be known at the BS while $\mathbf{H}$ is unknown over the entire transmission. Error in the estimation is defined by the channel estimation error matrix $\widetilde{\mathbf{H}} = [\widetilde{\mathbf{H}}_{1},\widetilde{\mathbf{H}}_{2},\ldots,\widetilde{\mathbf{H}}_{K}]$ in which each element of $\widetilde{\mathbf{H}}_k$ is an independent and identically distributed (i.i.d) complex Gaussian distribution variable with zero mean. Whereas, $\mathbb{E}\{\widetilde{\mathbf{H}}_{k}\widetilde{\mathbf{H}}_{k}^{H}\}=\mathbf{R}_{e,k}$ is the covariance matrix of the error matrix, independent of $\widehat{\mathbf{H}}_{k}$. Furthermore, the average CSIT error power is defined as $\sigma_{e,k}^2 \triangleq \mathbb{E}_{\widetilde{\mathbf{H}}_{k}}\big\{\norm{\widetilde{\mathbf{H}}_{k}}^2\big\} =\frac{1}{M}tr(\mathbf{R}_{e,k}) $. $\sigma_{e,k}^2$ is allowed to scale as $\mathcal{O}(P_{t}^{-\alpha})$ felicitating the  scaling of the CSIT quality with SNR, where $\alpha \in [0,\infty)$ is the quality scaling factor representing the quality of CSI at the BS in the high SNR regime  \cite{RS2016hamdi,NJindalMIMO2006,doppler2010Caire,DoF2013SYang}. Consequently, we write $\sigma_{e,k}^2= \mathcal{O}(P_{t}^{-\alpha})$ such that the error variance is assumed to scale exponentially with SNR. For $\alpha = \infty$, the average CSIT error power is equal to zero as $\sigma_{e,k}^2=0,\: \forall\, k\, \in \, \mathcal{K}$, resulting in a perfect CSIT scenario. On the other extreme, for $\alpha=0$, the CSIT quality remains invariant w.r.t SNR. Thus, a finite non-zero $\alpha$ leads to CSIT quality improvement as SNR increases, e.g., increasing the number of feedback bits with SNR. Here we truncate $\alpha \in [0,1]$. From a DoF perspective $\alpha=1$ corresponds to perfect CSIT \cite{RS2016hamdi}. 
\subsection{MIMO Rate Splitting}
Here we delineate the RS framework proposed for MIMO BC.
\subsubsection{Transmitter}
There are $Q_{k}\leq\min(M,Q)$ messages intended for user-$k$, $\forall k\in\mathcal{K}$, such that $\sum_{k=1}^{K}Q_{k}=Q_{p}=\min(M,KQ)$. These messages are expressed as $\mathbf{w}_k=\{ W_{1}^{k}, W_{2}^{k},\ldots, W_{Q_{k}}^{k}\}$, $k\in\mathcal{K}$. Each message of user-$k$ is split into a common part and a private part as $W_{i}^{k}=\{W_i^{c,k}, W_i^{p,k}\}$, $\forall i\in\{1,\ldots,Q_{k}\}$. The common parts $\mathbf{w}_{c,1},\ldots,\mathbf{w}_{c,K}$ of the messages of all users, with  $\mathbf{w}_{c,k}=\{W_{1}^{c,k}, W_{2}^{c,k},\ldots, W_{Q_{k}}^{c,k}\}$ denoting common parts of user-$k$, are combined into $Q_c,\,Q_{c}\in\{1,\ldots,\min(M,Q)\}$ common messages denoted by $\mathbf{w}_c\in\mathbb{C}^{Q_c\times1}$, and encoded together into a common stream vector of size $Q_c$ denoted by $\mathbf{s}_{c}=[s_{1}^{c},\ldots, s_{Q_c}^{c}]^{T}$. $\mathbf{s}_c$ will be decoded by all users. The private parts of user-$k$, $\mathbf{w}_{p,k}=\{W_{1}^{p,k},\ldots, W_{Q_{k}}^{p,k}\}\in\mathbb{C}^{Q_k\times1}
$ are independently encoded into a private stream vector $\mathbf{s}_{k} = [s_{1}^{p,k},\ldots, s_{Q_{k}}^{p,k}]^{T}$ meant to be decoded by the corresponding user-$k$ only. Therefore, the overall data stream vector to be transmitted is expressed as $\mathbf{s}=[\mathbf{s}_{c},\mathbf{s}_{1},..,\mathbf{s}_{K}]^T$. We use linear precoders $\mathbf{P}=[\mathbf{P}_{c},\mathbf{P}_{1},..,\mathbf{P}_{K}]$ to precode the data streams, where  $\mathbf{P}_{c}\in\mathbb{C}^{M\times Q_c}$ is the precoder for the common stream vector and $\mathbf{P}_{k}\in\mathbb{C}^{M\times Q_{k}}$ is the precoder for the private stream vector of user-$k$. The resulting transmit signal is $\mathbf{x} = \mathbf{P}\mathbf{s}$. The assumption is that $\mathbb{E}\{\mathbf{s}\mathbf{s}^H\}=\mathbf{I}$ thereby making the transmit power constraint as $\mathbb{E}\big\{tr(\mathbf{P}\mathbf{P}^{H})\big\} \leq P_t$. 
\subsubsection{MMSE Receiver and Rates}
\begin{figure}
\centering
\includegraphics[width=12cm,height=6.2cm]{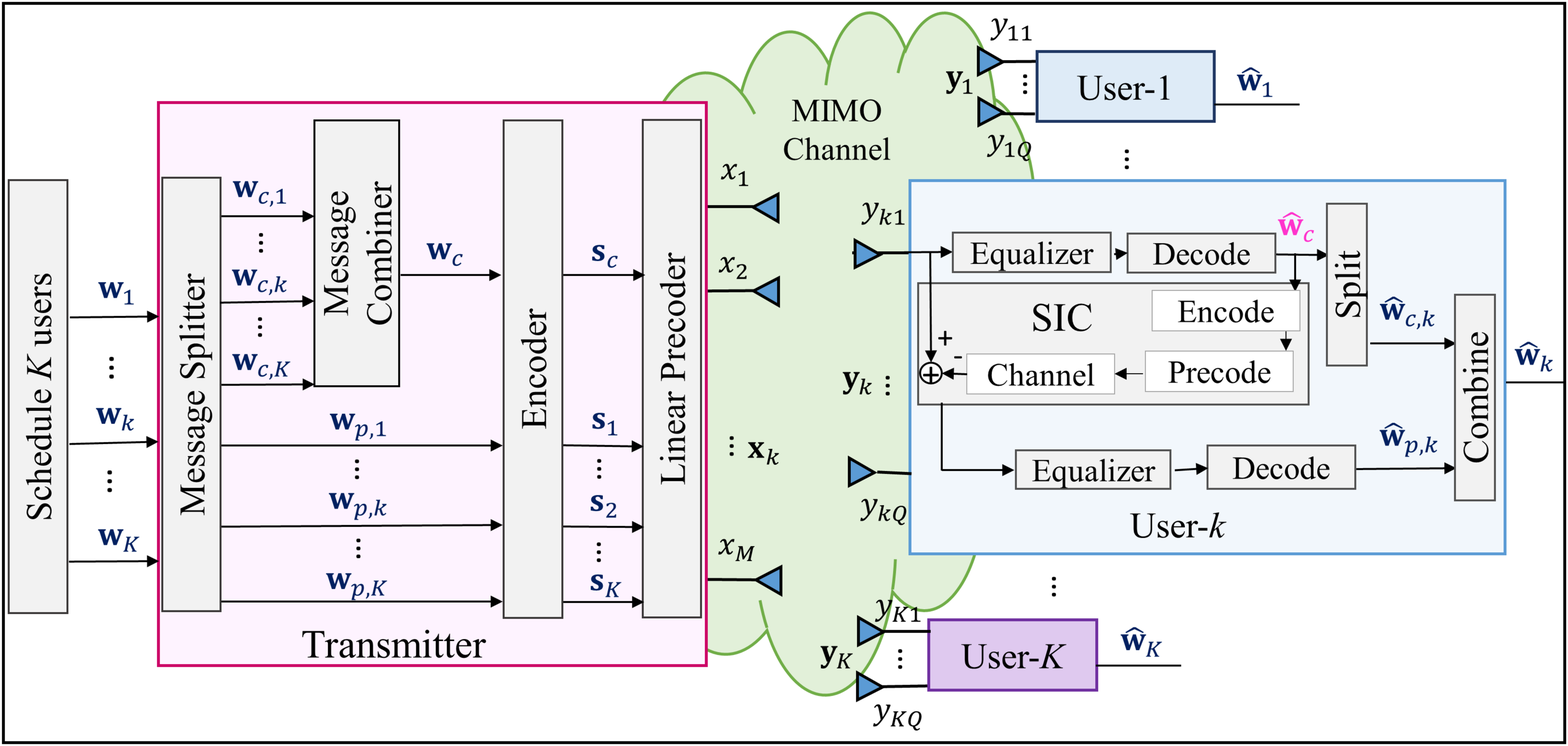}%
\caption{Transmission model of RS in MIMO BC}%
\label{fig:Txmodel}
\end{figure}

At user-$k$, first the common stream vector $\mathbf{s}_{c}$ is decoded into $\widehat{\mathbf{w}}_c$ by treating the interference from all private stream vectors as noise. Once the common stream vector is decoded and removed successfully using SIC, the private stream vector $\mathbf{s}_{k}$ of user-$k$ is decoded into $\widehat{\mathbf{w}}_{p,k}$ by treating interference from  private stream vectors of other users as noise. User-$k$ reconstructs its original message by extracting $\widehat{\mathbf{w}}_{c,k}$ from $\widehat{\mathbf{w}}_c$ and combining it with $\widehat{\mathbf{w}}_{p,k}$ to form $\widehat{\mathbf{w}}_{k}$. Fig.~\ref{fig:Txmodel} shows the $K$-user RS transmission model for MIMO BC. Next, we specify the instantaneous and ergodic rates for the common and private stream vectors (which are respectively denoted as common rate and private rate in the following). 

\par Since the precoder design at the BS is dependent on the channel estimate $\widehat{\mathbf{H}}$ while each user having perfect CSIR decodes its intended streams based on the exact channel $\mathbf{H}$, a joint fading state $\{\widehat{\mathbf{H}},\mathbf{H}\}$ determines the instantaneous common and private rates of each user. Assuming the signalling to be Gaussian, for a given channel realization, the instantaneous common and private rates $R_{z,k}(\mathbf{H},\widehat{\mathbf{H}}),\,z\in\{c,p\}$\footnote{To avoid redundancy, wherever possible, subscript ${z},\,z\in\{c,p\}$ will be used throughout the paper to simultaneously represent entities associated with the common and private stream vectors with $c$ representing entities associated with the common stream vector and $p$ with the private stream vector.} can be written as
\begin{equation}\label{eq:InstRate_PC}
\begin{split}
R_{c,k}(\mathbf{H},\widehat{\mathbf{H}})&=\log_{2}\det(\mathbf{I}+\mathbf{P}_{c}^{H}\mathbf{H}_{k}(\mathbf{R}_{c,k})^{-1}\mathbf{H}_{k}^{H}\mathbf{P}_{c}),\\
R_{p,k}(\mathbf{H},\widehat{\mathbf{H}})&=\log_{2}\det(\mathbf{I}+\mathbf{P}_{k}^{H}\mathbf{H}_{k}(\mathbf{R}_{p,k})^{-1}\mathbf{H}_{k}^{H}\mathbf{P}_{k}).
\end{split}
\end{equation}
The noise plus interference covariance matrices  $\mathbf{R}_{z,k}(\mathbf{H},\widehat{\mathbf{H}}),\,z\in\{c,p\}$ for the common and private stream vectors at user-$k$  are defined as
\begin{equation}\label{eq:a1}
\mathbf{R}_{c,k}(\mathbf{H},\widehat{\mathbf{H}}) = \mathbf{I}_{Q}+ \sum_{i=1}^{K} \mathbf{H}_{k}^H \mathbf{P}_{i} \mathbf{P}_{i}^H \mathbf{H}_{k},\;\;\mathbf{R}_{p,k}(\mathbf{H},\widehat{\mathbf{H}}) = \mathbf{I}_{Q}+ \sum_{i=1,i\neq k}^{K} \mathbf{H}_{k}^H \mathbf{P}_{i} \mathbf{P}_{i}^H \mathbf{H}_{k}.
\end{equation}
Denote the receive filters for common and private stream vectors at user-$k$ as $\mathbf{G}_{c,k}(\mathbf{H},\widehat{\mathbf{H}})\in\mathbb{C}^{Q_{c}\times Q}$ and $\mathbf{G}_{p,k}(\mathbf{H},\widehat{\mathbf{H}})\in\mathbb{C}^{Q_{k}\times Q}$, respectively. The estimated common stream vector is denoted as $\widehat{\mathbf{s}}_{c,k} = \mathbf{G}_{c,k}\mathbf{y}_{k}$. Assuming successful decoding and removal of the common stream vector, the private stream vector is estimated as $\widehat{\mathbf{s}}_k = \mathbf{G}_{p,k}\big(\mathbf{y}_k-\mathbf{H}_{k}^{H}\mathbf{P}_{c}\mathbf{s}_{c}\big)$. Therefore, Mean Square Error (MSE) matrices are written as
\begin{equation}\label{eq:MSE_matCP}
\mathbf{E}_{c,k}(\mathbf{H},\widehat{\mathbf{H}})=\mathbb{E}\big[(\widehat{\mathbf{s}}_{c,k}-\mathbf{s}_c)(\widehat{\mathbf{s}}_{c,k}-\mathbf{s}_c)^H\big],\;\; \mathbf{E}_{p,k}(\mathbf{H},\widehat{\mathbf{H}})=\mathbb{E}\big[(\widehat{\mathbf{s}}_{k}-\mathbf{s}_k)(\widehat{\mathbf{s}}_{k}-\mathbf{s}_k)^H\big].
\end{equation}
Minimizing the MSEs by solving $\frac{\partial \mathbf{E}_{z,k}}{\partial \mathbf{G}_{z,k}}=0,\, z\in\{c,p\}$ leads to respective Minimum MSE (MMSE) filters
\begin{equation}\label{eq:a3}
\begin{split}
\mathbf{G}_{c,k}^{\textrm{MMSE}}(\mathbf{H},\widehat{\mathbf{H}})= \arg\min_{\mathbf{G}_{c,k}} \mathbb{E}[\|\mathbf{G}_{c,k}\mathbf{y}_{k}-\mathbf{s}_{c}\|^{2}]=\mathbf{P}_{c}^{H}\mathbf{H}_{k}(\mathbf{H}_{k}^H\mathbf{P}_{c}\mathbf{P}_{c}^{H}\mathbf{H}_{k} + \mathbf{R}_{c,k})^{-1},
\end{split}
\end{equation}
\begin{equation}\label{eq:a4}
\begin{split}
\mathbf{G}_{p,k}^{\textrm{MMSE}}(\mathbf{H},\widehat{\mathbf{H}})= \arg\min_{\mathbf{G}_{p,k}} \mathbb{E}[\|\mathbf{G}_{p,k}(\mathbf{y}_{k}-\mathbf{H}_{k}^H\mathbf{P}_{c}\mathbf{s}_{c})-\mathbf{s}_{k}\|^{2}] =\mathbf{P}_{k}^{H}\mathbf{H}_{k}(\mathbf{H}_{k}^H\mathbf{P}_{k}\mathbf{P}_{k}^{H}\mathbf{H}_{k} + \mathbf{R}_{p,k})^{-1}.
\end{split}
\end{equation}
Substituting the MMSE filters (\ref{eq:a3}) and (\ref{eq:a4}) in (\ref{eq:MSE_matCP}), respectively, the MMSE matrices for common and private stream vectors are calculated as 
\begin{equation} \label{eq:a5}
\mathbf{E}_{c,k}^{\textrm{MMSE}}(\mathbf{H},\widehat{\mathbf{H}})=(\mathbf{I}+\mathbf{P}_{c}^{H}\mathbf{H}_{k}(\mathbf{R}_{c,k})^{-1}\mathbf{H}_{k}^{H}\mathbf{P}_{c})^{-1},\;\;
\mathbf{E}_{p,k}^{\textrm{MMSE}}(\mathbf{H},\widehat{\mathbf{H}})=(\mathbf{I}+\mathbf{P}_{k}^{H}\mathbf{H}_{k}(\mathbf{R}_{p,k})^{-1}\mathbf{H}_{k}^{H}\mathbf{P}_{k})^{-1}.
\end{equation}
Using MMSE matrices in (\ref{eq:a5}) and equation (\ref{eq:InstRate_PC}), we obtain the instantaneous rate expressions as $R_{z,k}(\mathbf{H},\widehat{\mathbf{H}})=\log_2{\det{\big(\mathbf{E}_{z,k}^{\textrm{MMSE}}(\mathbf{H},\widehat{\mathbf{H}})\big)^{-1}}},\,z\in\{c,p\}.$ 
Partial knowledge of the CSI at the BS may result in overestimation of the instantaneous rates, rendering them unachievable \cite{RS2016hamdi}. A robust method would be to design the precoders based on the ER assuming that the transmission is delay unlimited. The ERs for the common and private stream vectors of user-$k$ are defined as 
\begin{equation}\label{eq:ER_c}
\bar{R}_{z,k}\triangleq \mathbb{E}_{\{\mathbf{H},\widehat{\mathbf{H}}\}}\{R_{z,k}(\mathbf{H},\widehat{\mathbf{H}})\},\,z\in\{c,p\}.
\end{equation}
ER characterizes the long-term performance of user-$k$ over all possible joint fading states. To ensure that each user is able to successfully decode the common stream vector, it needs to be sent at an ER $\bar{R}_c=\min_{j}\big\{\bar{R}_{c,j}\big\}_{j=1}^{j=K}$.
The common rate is shared by all users with $\bar{C}_k$ denoting the share allocated to user-$k$ such that $\sum_{k\in\mathcal{K}}\bar{C}_{k}=\bar{R}_c$. Therefore, the total ER achieved by user-$k$ is equal to $\bar{R}_{k,tot}=\bar{R}_{p,k}+\bar{C}_k$.
\subsection{Two User Example}
To better illustrate RS, we consider a two-user case. At the BS, $Q_{k},\:k\in\{1,2\}$ messages of both users denoted by $ \mathbf{w}_1=\{W_{1}^{1},\ldots, W_{Q_{1}}^{1}\}$ and $ \mathbf{w}_2=\{W_{1}^{2},\ldots, W_{Q_{2}}^{2}\}$ are respectively split into sub-messages as $  \mathbf{w}_1=\big\{\{W_{1}^{c,1},W_{1}^{p,1}\},\ldots, \{W_{Q_{1}}^{c,1},W_{Q_{1}}^{p,1}\}\big\}$ and $  \mathbf{w}_2=\big\{\{W_{1}^{c,2},W_{1}^{p,2}\},\ldots, \{W_{Q_{2}}^{c,2},W_{Q_{2}}^{p,2}\}\big\}$. Sub-messages $\{\{W_{1}^{c,k},\ldots,W_{Q_{k}}^{c,k}\}\mid\forall k\in\,\{1,2\}\}$ are combined and jointly encoded into a common stream vector $\mathbf{s}_c$ of size $Q_{c}$. Whereas, the private sub-messages $\{W_{1}^{p,1},\ldots,W_{Q_{1}}^{p,1}\}$ and $\{W_{1}^{p,2},\ldots,W_{Q_{2}}^{p,2}\}$ are respectively encoded into streams $\mathbf{s}_1$ of size $Q_{1}$ and $\mathbf{s}_2$ of size $Q_{2}$. For instance, in our simulations, we consider $M=4$, $Q=2$, $Q_{1}=Q_{2}=2$ and $Q_{c}=2$. The transmit signal is formed by precoding and superposing the encoded data stream vectors $\mathbf{s}_c$, $\mathbf{s}_1$, $\mathbf{s}_2$, which is expressed as $\mathbf{x}=\mathbf{P}_{c}\mathbf{s}_{c} + \mathbf{P}_{1}\mathbf{s}_{1} + \mathbf{P}_{2}\mathbf{s}_{2} $. At the user side, both users decode their intended streams using SIC. At user-$1$, first $\mathbf{s}_c$ is decoded by treating $\mathbf{s}_1$ and $\mathbf{s}_2$ as noise. Assuming $\mathbf{s}_c$ is successfully decoded, user-$1$ then removes it from the received signal and decodes $\mathbf{s}_1$ by treating $\mathbf{s}_2$ as noise. Similarly, user-$2$ decodes its intended common and private streams sequentially. The two-user RS case reduces to the conventional MU--MIMO strategy by simply turning off the common streams, i.e., allocating no power to the common stream vector. Considering the other extreme of fully decoding the interference, we look at the MIMO NOMA strategy. By encoding the entire $\mathbf{w}_2$ into $\mathbf{s}_c$, allocating no power to $\mathbf{s}_2$ and encoding $\mathbf{w}_1$ into $\mathbf{s}_1$, the two-user RS reduces to MIMO NOMA with decoding order $1\rightarrow2$, where the message of user-$2$ is decoded by both users and the message of user-$1$ is decoded by user-$1$ only. Table \ref{tab:Table_Map} illustrates the mapping of messages to streams.
\begin{table}[H]
\caption{Messages mapped to streams in different schemes.}
    \centering
    \begin{tabular}{L{2cm}  C{2cm} C{2cm} C{2cm}}
 \hline
  & $\mathbf{s}_1$ & $\mathbf{s}_2$ & $\mathbf{s}_c$\\
 \hline
 RS & $\mathbf{w}_{p,1}$ &$\mathbf{w}_{p,2}$& $\mathbf{w}_{c,1}, \mathbf{w}_{c,2}$\\ 
 MU-MIMO & $\mathbf{w}_1$ & $\mathbf{w}_2$ & -\\ 
 MIMO NOMA & $\mathbf{w}_1$ & - & $\mathbf{w}_2$\\ 
 \hline
  & \multicolumn{2}{c}{decoded by its corresponding user} & decoded by  \\
 & \multicolumn{2}{c}{and treated as noise by the other user} & both users\\
\end{tabular}
    \label{tab:Table_Map}
\end{table}
\section{Problem Formulation}\label{ProbForm}
In this section, our objective is to formulate the precoder optimization problem for the proposed system model. A naive approach for precoder design is to assume the channel estimate $\widehat{\mathbf{H}}$ to be perfect and optimize the instantaneous precoder $\mathbf{P}$ by maximizing the instantaneous WSR subject to the instantaneous power constraint $tr(\mathbf{P}\mathbf{P}^{H})\leq {P}_t$. However, this approach might not be able to cope with MU interference and it may lead the BS to transmit at undecodable rates \cite{RS2016hamdi}. A more robust approach to designing precoders is to maximize the Weighted Ergodic Sum-Rate (WESR) which captures the long-term WSR performance of all users to ensure reliable transmission. We first define the WESR for RS as $\bar{R}_{RS}(\boldsymbol{\mu})=\sum_{k=1}^{K}\mu_{k}\bar{R}_{k,tot}$, where $\mu_k$ is the weight allocated to user-$k$ and $\boldsymbol{\mu}=\{\mu_1,...,\mu_K\}$.
Next, we consider the Weighted Average SR (WASR) optimization approach to maximize WESR at the BS with imperfect instantaneous CSIT. Though it is difficult to predict the instantaneous rates at the BS, the BS can instead access the Average-Rate (AR).
\theoremstyle{plain}
\begin{definition}
For a given channel state estimate $\widehat{\mathbf{H}}$ and precoder $\mathbf{P}(\widehat{\mathbf{H}})$, AR is defined as the expected performance over the CSIT error distribution. The ARs for the common and private stream vectors at user-$k$ are given by 
\begin{equation}\label{eq:AR_c}
\widehat{R}_{z,k}(\mathrm{\widehat{\mathbf{H}}})=\mathbb{E}_{\mathrm{\mathbf{H}}|\mathrm{\widehat{\mathbf{H}}}}\big\{R_{z,k}(\mathbf{H},\widehat{\mathbf{H}})|\mathrm{\widehat{\mathbf{H}}}\big\},\,z\in\{c,p\}.
\end{equation}
\end{definition}
AR should not be confused with ER. While ER captures the long-term performance over all channel states, AR measures the short-term expected performance over CSIT error distribution for one channel estimate. By using the law of total expectation and the definition of AR, the relation between ER and AR of the common and private stream vectors at user-$k$ is established as $\bar{R}_{z,k}=\mathbb{E}_{\mathrm{\widehat{\mathbf{H}}}}\big\{\widehat{R}_{z,k}(\widehat{\mathbf{H}})\big\},\,z\in\{c,p\}$\cite{RS2016hamdi}.
The share of the AR allocated to user-$k$ corresponding to the common stream vector is defined as $\widehat{C}_k$ such that $\sum_{k=1}^{K} \widehat{C}_k=\widehat{R}_{c}$ and
$\widehat{R}_c$ must not be greater than $\min_{k\in\mathcal{K}} \{\widehat{R}_{c,k}\}$. For calculating AR of the common stream vector $\widehat{R}_c$, we write
\begin{equation}\label{eq:min_exp}
\min_{k \in \mathcal{K}}\big\{\mathbb{E}_{\widehat{\mathbf{H}}}\{\widehat{R}_{c,k}\}\big\}\geq\mathbb{E}_{\widehat{\mathbf{H}}}\big\{\min_{k\in\mathcal{K}} \{\widehat{R}_{c,k}\}\big\},
\end{equation}
as interchanging minimization and expectation in (\ref{eq:min_exp}) does not increase the value of left hand side. Consequently, following Law of Large Numbers (LLN) we approximate the ER of each stream by averaging its AR over all channel states and thereby remove dependencies among channel states. This allows us to  decompose the WESR maximization problem with short term\footnote{For tractability, we replace the long-term power constraint   $\mathbb{E}_{\mathbf{H}}\{tr(\mathbf{P}\mathbf{P}^{H})\}\leq P_{t}$ with short-term power constraints \cite{RS2016hamdi}.} power constraints to a WASR maximization problem for each $\widehat{\mathbf{H}}$ defined as
\begin{subequations}\label{eq:3}
\begin{align}
\widehat{R}_{RS}(P_{t},\boldsymbol{\mu})&=\max_{\mathbf{P,\widehat{c}}} \sum_{k=1}^{K} \mu_{k} \widehat{R}_{k,tot} \\ 
& \widehat{C}_{1} + \widehat{C}_{2} + \ldots \widehat{C}_{K} \leq \widehat{R}_{c}\\
& tr(\mathbf{P}\mathbf{P}^H) \leq {P_{t}}\\
& \mathbf{\widehat{c}}\geq \mathbf{0},
\end{align}
\end{subequations}
where ${\mathbf{\widehat{c}}}=[\widehat{C}_{1},\widehat{C}_{2},\ldots,\widehat{C}_{K}]$ is the average common rate vector and $\widehat{R}_{k,tot}= \widehat{R}_{p,k} + \widehat{C}_{k}$.
After formulating the WASR problem for the RS scheme, we look at the effect of the instantaneous CSIT quality on its long-term performance and observe how the RS scheme fares against conventional multiple access schemes. We do that by looking at the sum-DoF analysis. 
\subsection{Common message and DoF Analysis}
DoF is the total number of interference free streams that can be transmitted simultaneously in a single channel use \cite{RSintro16bruno}. The sum-DoF for RS is defined as 
\begin{equation}\label{eq:b4}
 d_{s}^{\textrm{RS}}=\lim_{P_t \rightarrow \infty} \frac{\mathbb{E}_{\mathrm{\widehat{\mathbf{H}}}}\{\widehat{R}_{RS}(P_t)\}}{\log_{2}(P_t)},
\end{equation}
where $\widehat{R}_{RS}(P_t)$ is the Average SR (ASR)
and is equal to $\widehat{R}_{RS}(P_t,\boldsymbol{\mu})$ for equal user weights, i.e., $\mu_{k}=1,\,\forall k\in\mathcal{K}$. We aim at establishing the sum-DoF achieved by the RS scheme for symmetric MIMO BC transmission with imperfect CSIT under the assumption that the channel $\widehat{\mathbf{H}}_{k}$ is full rank and CSIT error matrix $\widetilde{\mathbf{H}}_{k}$ is isotropically distributed. It should be noted that these assumptions are not necessary for optimization. With $Q$ receive antennas at each user, $Q_p$ being the total number of private streams transmitted and $Q_c$ as the number of common streams, the sum-DoF achieved by the RS precoding scheme is
\begin{equation}\label{eq:4}
   d_{s}^{\textrm{RS}}:\begin{dcases}
   Q_c(1-\alpha)+Q_{p}\alpha,& M\in \{2Q, 3Q,\ldots, KQ\}\\ 
   M,&M\leq Q.
  \end{dcases}
\end{equation}
For comparison, we consider the conventional MU--MIMO and MIMO NOMA schemes with sum-DoF expressed as 
\begin{equation}\label{eq:5}
 d_{s}^{\textrm{MU-MIMO}}= \max\big(\min(M,Q),Q_{p}\alpha\big).
\end{equation}
\begin{equation}\label{eq:nomadof}
 d_{s}^{\textrm{NOMA}}=\min(M,Q).
\end{equation}
The procedure to obtain the sum-DoF achieved by all schemes is relegated to Appendix A. Following the principle of SC-SIC, the ASR achieved by MIMO NOMA (SC--SIC) is limited by the decodability of all users messages decoded in the last place. This restricts the sum-DoF to a maximum value of $\min(M,Q)$. For MU--MIMO, the sum-DoF can attain a maximum value of $Q_{p}$ when $\alpha=1$. As $\alpha$ falls below $1$, detrimental effects of interference lead to a decrease in its sum-DoF. Once $\alpha$ goes further down, CSIT quality deteriorates to a point where it is not conducive enough to support MU transmission and transmitting to a single user yields a better sum-DoF. However, in the RS scheme, the presence of common messages allows the transmitter to adjust the power allocated to the private stream vectors in a way that the interference is always at the level of noise. Thus, the DoF of the private stream vectors is maintained at $Q_{p}\alpha$ by scaling down the power allocated to the private stream vectors
to $\mathcal{O}(P_t^{\alpha})$. The remaining power which scales as $\mathcal{O}(P_t)$ is allocated to the common stream vector. The DoF gain achieved by the common streams is $Q_c(1-\alpha)$. For $\alpha\, \in(0,1)$ and $Q_{c}=\min(M,Q)$, the sum-DoF of RS is strictly greater than the sum-DoF of both MU--MIMO and MIMO NOMA. 
\par Though optimization does not improve the achievable sum-DoF, it does play a significant role in improving the rate performance. As $\widehat{R}_{\textrm{MU-MIMO}}(P_t)$ can be obtained by switching off the common streams, the inequality $\widehat{R}_{\textrm{RS}}(P_t)\geq \widehat{R}_{\textrm{MU-MIMO}}(P_t)$ is guaranteed for the entire range of SNR.
The results in Section \ref{NumRes} validate the theoretical assertions.
\section{Optimization Framework}\label{OptFrame}
The optimization problem and sub-problems of (\ref{eq:3}) are non-deterministic in nature and thus solving them becomes very difficult in their current form. We adopt a three-step approach to solve the optimization problem (\ref{eq:3}). First, we use the Sample Average Approximation (SAA) method to obtain a deterministic approximation of the problem, then we transform the WASR problem to a part-wise convex Weighted Average MMSE (WAMMSE) problem making it solvable. Finally, we use vectorization to reduce the matrix variables to their vectorized forms, thereby making WAMMSE problem tractable to solve. Using AO, we obtain the precoders and consequently, the optimized rate for a given weight vector $\boldsymbol{\mu}$.. 
\subsection{Sample Average Approximation}
We first consider a set of $N$ i.i.d channel samples indexed $\mathcal{N}=\{1,2,\ldots,N\}$ drawn from a distribution with density $f_{\mathrm{\mathbf{H}}|\mathrm{\widehat{\mathbf{H}}}}\big\{\mathbf{H}|\mathbf{\widehat{H}}\big\}$. Therefore, for a given channel estimate $\widehat{\mathbf{H}}$ we have $N$ channel samples denoted as $\mathbb{H}^{(N)} \triangleq \big\{ \mathbf{H}^{(n)}=\widehat{\mathbf{H}}+\widetilde{\mathbf{H}}^{(n)}\mid \widehat{\mathbf{H}}, n \in \mathcal{N} \big\}$.
Using the channel realizations and Sample Average Functions (SAFs) defined as: $\widehat{R}_{z,k}^{(N)} \triangleq \frac{1}{N}\sum_{n=1}^{N}R_{z,k}^{(n)},\,z\in \{c,p\}$,  we approximate the average rates. Here, $R_{z,k}^{(n)}\triangleq R_{z,k}(\mathbf{H}^{(n)}),\,z\in \{c,p\}$ are the common and private rates associated with the $n^{th}$ realization. The SAA of problem (\ref{eq:3}) is
\begin{subequations}\label{eq:7}
\begin{align}
\widehat{R}_{RS}^{(N)}(P_{t},\boldsymbol{\mu}) &= \max_{\mathbf{P},\widehat{\mathbf{c}}} \sum_{k=1}^{K} \mu_{k} \widehat{R}_{k,tot}^{(N)} \\ 
& \widehat{C}_{1} + \widehat{C}_{2} + \ldots \widehat{C}_{K} \leq \widehat{R}_{c}^{(N)}\\
& tr(\mathbf{P}\mathbf{P}^H) \leq {P_{t}}\\
& \widehat{\mathbf{c}}\geq \mathbf{0},
\end{align}
\end{subequations}
where $\widehat{R}_{k,tot}^{(N)}=\widehat{R}_{p,k}^{(N)} + \widehat{C}_{k}$ and $\widehat{R}_{c}^{(N)}\triangleq \min_{j}\{\widehat{R}_{c,j}^{(N)}\}_{j=1}^{K}$.
The rates obtained here are bounded \cite{RS2016hamdi} and therefore by applying LLN, it can be inferred that 
\begin{equation}\label{eq:8}
\lim_{N \rightarrow \infty} \widehat{R}_{z,k}^{(N)}(\mathbf{P})= \widehat{R}_{z,k}(\mathbf{P}), \forall \mathbf{P} \in \mathbb{P},\,z\in\{c,p\}.
\end{equation}
The set $\mathbb{P}$ defined as $\{tr(\mathbf{P}\mathbf{P}^{H}) \leq P_t\mid\mathbf{P}\in \mathbb{P}\}$ is the feasible set of precoders for which the rate functions are bounded, continuous and differentiable in $\mathbf{P}$, thereby making convergence in (\ref{eq:8}) uniform in $\mathbf{P}$.
The ARs are also continuous and differentiable \cite{RS2016hamdi} and therefore using  (\ref{eq:8}) we obtain
\begin{equation}\label{eq:a18}
\lim_{N\rightarrow \infty}\widehat{R}_{k,tot}^{(N)}=\widehat{R}_{k,tot}\;\;\; \forall\;\mathbf{P}\in\mathbb{P}.
\end{equation}
Based on (\ref{eq:8}), (\ref{eq:a18}), we obtain that as $N\rightarrow \infty$, the optimum solutions of the SAA in problem (\ref{eq:7}) converges to the solution of the stochastic problem in (\ref{eq:3})\cite{Conv_opt,RS2016hamdi}.
\subsection{WASR $\rightarrow$ WAMMSE}
In this subsection, we aim at solving the sample average approximated problem (\ref{eq:7}) by using the methods adopted in \cite{wmmse08} to transform problem (\ref{eq:7}) into an equivalent WAMMSE form. 
\par First we define Augmented Weighted Mean Square Error (AWMSE) for common and private stream vectors as
\begin{equation}\label{eq:9}
\xi_{z,k}(\mathbf{H},\widehat{\mathbf{H}})=tr(\mathbf{U}_{z,k}\mathbf{E}_{z,k})-\log{\det{(\mathbf{U}_{z,k})}},\,z\in \{c,p\}.
\end{equation}
where $\mathbf{U}_{z,k},\, z\in \{c,p\}$ are instantaneous weights introduced for common and private MSE matrices of user-$k$. Next we aim to establish the Rate-WMMSE relationship by optimizing the AWMSEs w.r.t equalizers (filters) and weights. By solving $\frac{\partial \xi_{z,k}(\mathbf{H},\widehat{\mathbf{H}})}{\partial \mathbf{G}_{z,k}}=0$, the optimum equalizers are obtained as $\mathbf{G}_{z,k}^{*}=\mathbf{G}_{z,k}^{\mathrm{MMSE}}$, $z\in \{c,p\}$. Substituting the optimum equalizers in (\ref{eq:9}), we get
\begin{equation}\label{eq:10}
\xi_{z,k}\big(\mathbf{G}_{z,k}^{\mathrm{MMSE}}\big)=tr(\mathbf{U}_{z,k}\mathbf{E}_{z,k}^{\mathrm{MMSE}})-\log{\det{(\mathbf{U}_{z,k})}},\,z\in \{c,p\}.
\end{equation}
By solving $\frac{\partial \xi_{z,k}\big(\mathbf{G}_{z,k}^{\mathrm{MMSE}}\big)}{\partial \mathbf{U}_{z,k}}=0$, the optimum MMSE weights are obtained as $\mathbf{U}_{z,k}^{*}=\mathbf{U}_{z,k}^{\mathrm{MMSE}}\triangleq (\mathbf{E}_{z,k}^{\mathrm{MMSE}})^{-1}$, $z\in \{c,p\}$. Subsituting the obtained optimum weights for the weights in equation (\ref{eq:10}), the instantaneous Rate-WMMSE relationship is established as 
\begin{equation}\label{eq:11}
\xi_{c,k}(\mathbf{H},\widehat{\mathbf{H}})\triangleq \min_{\mathbf{G}_{c,k},\mathbf{U}_{c,k}}\xi_{c,k}=Q_c-R_{c,k}(\mathbf{H},\widehat{\mathbf{H}}),\;\;
\xi_{p,k}(\mathbf{H},\widehat{\mathbf{H}})\triangleq \min_{\mathbf{G}_{p,k},\mathbf{U}_{p,k}}\xi_{p,k}=Q_{k}-R_{p,k}(\mathbf{H},\widehat{\mathbf{H}}).
\end{equation}
Based on the principle of SAA, the AR-WAMMSE relationship is derived by taking the expectation over the conditional distribution of channel $\mathbf{H}$ for a given channel estimate $\widehat{\mathbf{H}}$ and is written as
\begin{equation}\label{eq:a11}
\widehat{\xi}_{c,k}\triangleq \mathbb{E}_{\mathrm{\mathbf{H}}|\mathrm{\widehat{\mathbf{H}}}}\big\{\min_{\mathbf{G}_{c,k},\mathbf{U}_{c,k}}\xi_{c,k} | \widehat{\mathbf{H}}\big\}= Q_c-\widehat{R}_{c,k},\;\;
\widehat{\xi}_{p,k}\triangleq \mathbb{E}_{\mathrm{\mathbf{H}}|\mathrm{\widehat{\mathbf{H}}}}\big\{\min_{\mathbf{G}_{p,k},\mathbf{U}_{p,k}}\xi_{p,k} | \widehat{\mathbf{H}}\big\}= Q_{k}-\widehat{R}_{p,k}.
\end{equation}
\par Next, we use the SAFs to obtain the deterministic equivalent relations of (\ref{eq:a11}). Taking the $N$ i.i.d channel samples, the average AWMSEs are $\widehat{\xi}_{z,k}^{(N)}\triangleq \frac{1}{N}\sum_{n=1}^{N}{\xi}_{z,k}^{(n)}$, $z\in \{c,p\}$, where ${\xi}_{z,k}^{(n)}$, $\mathbf{G}_{z,k}^{(n)}$, $\mathbf{U}_{z,k}^{(n)}$, $z\in \{c,p\}$ are all associated with the $n^{th}$ realization in $\mathbb{H}^{(N)}$.
For ease of notation, we use $\mathbf{G}$ to represent the set of equalizers for common and private stream vectors, i.e., $\mathbf{G} \triangleq \{\mathbf{G}_{z,k} \mid \forall \, k \in \mathcal{K}\}$, where $\mathbf{G}_{z,k}\triangleq \{\mathbf{G}_{z,k}^{(n)}\mid \forall \, n \in \mathcal{N},\,z\in\{c,p\} \}$. Following the same method, we obtain the set of weights for common and private streams, denoted as  $\mathbf{U}$.
\par Following the LLN as in (\ref{eq:8}) and the approach used to obtain (\ref{eq:11}), the AR-WAMMSE  in (\ref{eq:a11}) is written as 
\begin{equation}\label{eq:12}
\big(\widehat{\xi}_{c,k}^{\;\mathrm{MMSE}}\big)^{(N)} \triangleq \min_{\mathbf{G}_{c,k},\mathbf{U}_{c,k}}\widehat{\xi}_{c,k}^{(N)}=Q_c-\widehat{R}_{c,k}^{(N)},\;\;  
\big(\widehat{\xi}_{p,k}^{\;\mathrm{MMSE}}\big)^{(N)} \triangleq \min_{\mathbf{G}_{p,k},\mathbf{U}_{p,k}}\widehat{\xi}_{p,k}^{(N)}=Q_{k}-\widehat{R}_{p,k}^{(N)}.
\end{equation}
The sets of optimum MMSE equalizers and weights associated with equation (\ref{eq:12}) are defined as $\mathbf{G}^{\mathrm{MMSE}} \triangleq \big\{\big(\mathbf{G}_{z,k}^{\mathrm{MMSE}}\big)^{(n)}\mid z\in \{c,p\}, \forall\,n\in \mathcal{N},\,\forall k\in \mathcal{K} \big\}$, $\mathbf{U}^{\mathrm{MMSE}} \triangleq \big\{\big(\mathbf{U}_{z,k}^{\mathrm{MMSE}}\big)^{(n)}\mid z\in \{c,p\}, \forall\,n\in \mathcal{N},\,\forall k\in \mathcal{K}\big\}$ respectively. 
Motivated by the AR-WAMMSE relationship in (\ref{eq:12}), the deterministic WAMMSE optimization problem is formulated as
\begin{subequations}\label{eq:13}
\begin{align}
&\min_{\mathbf{P},\widehat{\mathbf{x}},\mathbf{U},\mathbf{G}} \sum_{k=1}^{K} \mu_{k} \widehat{\xi}_{k,tot}^{(N)} \\ 
& \widehat{X}_{1} + \widehat{X}_{2} + \ldots \widehat{X}_{K}+Q_c \geq \widehat{\xi}_{c}^{(N)}\\
& tr(\mathbf{P}\mathbf{P}^H) \leq {P_{t}}\\
& \widehat{\mathbf{x}}\leq \mathbf{0},
\end{align}
\end{subequations}
where $\widehat{\xi}_{c}^{(N)}=\max\{\widehat{\xi}_{c,k}^{(N)}\}_{k=1}^{K}$, $\widehat{\xi}_{k,tot}^{(N)}=\widehat{\xi}_{p,k}^{(N)} + \widehat{X}_{k} $ and $\widehat{\mathbf{x}} = \{\widehat{X}_{1}, \widehat{X}_{2},\ldots,\widehat{X}_{K}\} = - \widehat{\mathbf{c}}$. (\ref{eq:13}) is optimized w.r.t $(\mathbf{U},\mathbf{G})$ by minimizing individual AWMSEs shown in (\ref{eq:12}) as the AWMSEs are decoupled in their corresponding weights and equalizers. This can be validated by showing that for a given precoder $\mathbf{P}$, the KKT optimality conditions of (\ref{eq:13}) are satisfied by the MMSE solution $\{\mathbf{U}^{\mathrm{MMSE}},\mathbf{G}^{\mathrm{MMSE}}\}$. Consequently, it can be shown that for the MMSE solution, (\ref{eq:13}) boils down to (\ref{eq:7}) and the Rate-WMMSE relationship is not just limited to the global optimum solution but can be extended to the entire set of stationary points. For any point $\{\mathbf{U}^{*},\mathbf{G}^{*},\mathbf{P}^{*},\widehat{\mathbf{x}}^{*}\}$ satisfying the KKT optimality conditions of (\ref{eq:13}), also satisfies the KKT optimality conditions of (\ref{eq:7}), with $\widehat{\mathbf{c}}=-\widehat{\mathbf{x}}$  \cite{RS2016hamdi}. Therefore, as $N\rightarrow \infty$, solving (\ref{eq:13}) yields a solution for (\ref{eq:7}), which in turn, converges to a solution of the WASR problem in (\ref{eq:3}).
\subsection{Vectorization and Alternate Optimization}
Problem (\ref{eq:13}) is non-convex for the joint optimization of variables $\mathbf{G},\mathbf{U}$, $\widehat{\mathbf{x}}$ and $\mathbf{P}$ but it is convex for each block of variables if the other two are fixed. Therefore, we utilize the AO algorithm with 2 steps, 1) updating the equalizers $\mathbf{G}$ and weights $\mathbf{U}$ by using (fixing) the precoders $\mathbf{P}$ from the previous iteration and 2) updating the precoders $\mathbf{P}$ and the message split $\mathbf{x}$ by solving the optimization problem for a given $\mathbf{G}$ and  $\mathbf{U}$. Unlike the MISO case in \cite{mao2017rate}, optimization in (\ref{eq:13}) encounters difficulties of optimizing matrix variables. Furthermore, the presence of $\mathbf{R}_{z,k}^{-1},\,z\in\{c,p\}$ matrices in the AWMSE expressions makes the optimization intractable. Bearing that in mind, we first use vectorization and deduce the objective function into a tractable form. Let us consider the augmented AWMSE expression (\ref{eq:9}) for the common stream. In step 2 of the AO, the weights are fixed. However, the calculation of the term $tr(\mathbf{U}_{c,k}\mathbf{E}_{c,k})$ introduces difficulties due to the aforementioned reasons and thus we try to simplify the expression and consequently the entire objective function into a solvable form. Expanding $tr(\mathbf{U}_{c,k}\mathbf{E}_{c,k})$ it follows,
\begin{equation}\label{eq:14}
\begin{split}
tr(\mathbf{U}_{c,k}\mathbf{E}_{c,k}) &= tr\Big(\mathbf{U}_{c,k}\mathbb{E}\big\{(\mathbf{G}_{c,k}\mathbf{y}_{k}-\mathbf{s}_{c})(\mathbf{G}_{c,k}\mathbf{y}_{k}-\mathbf{s}_{c})^{H}\big\}\Big)= tr\Big(\mathbf{U}_{c,k}\big \{ \mathbf{G}_{c,k}\mathbf{H}_{k}^{H}\mathbf{P}_{c}\mathbf{P}_{c}^{H}\mathbf{H}_{k}{\mathbf{G}_{c,k}^H} \\
&+\mathbf{G}_{c,k}\mathbf{H}_{k}^{H}(\sum_{i=1}^{K}\mathbf{P}_{i} \mathbf{P}_{i}^H)\mathbf{H}_{k}{\mathbf{G}_{c,k}^H}-\mathbf{G}_{c,k}\mathbf{H}_{k}^{H}\mathbf{P}_{c}- \mathbf{P}_{c}^{H}\mathbf{H}_{k}{\mathbf{G}_{c,k}^H}+\sigma_{n}^{2}\mathbf{G}_{c,k}{\mathbf{G}_{c,k}^H}+\mathbf{I} \big\}\Big).
\end{split}
\end{equation}
Simplifying\footnote{Applying matrix manipulation $tr(\mathbf{AB})=tr(\mathbf{BA})$ and  $tr(\mathbf{ABC})=vec(\mathbf{A}^{H})^{H}(\mathbf{I}\otimes \mathbf{B})vec(\mathbf{C})$ we transform (\ref{eq:14}) to (\ref{eq:a14}).} the expanded expression in (\ref{eq:14}), we get
\begin{equation}\label{eq:a14}
\begin{split}
tr(\mathbf{U}_{c,k}\mathbf{E}_{c,k})-\log\det(\mathbf{U}_{c,k})= \mathbf{p}_{c}^{H}\mathbf{A}_{c,k}^{'}\mathbf{p}_{c}+\sum_{i=1}^{K}\mathbf{p}_{i}^{H}\mathbf{A}_{c,k}\mathbf{p}_{i} - \mathbf{a}_{c,k}^{H}\mathbf{p}_{c}- \mathbf{p}_{c}^{H}\mathbf{a}_{c,k}+{\Phi}_{c,k}. 
\end{split}
\end{equation}
Similarly, the resultant expression for $tr(\mathbf{U}_{p,k}\mathbf{E}_{p,k})$ is  
\begin{equation}\label{eq:15}
\begin{split}
tr(\mathbf{U}_{p,k}\mathbf{E}_{p,k})-\log\det(\mathbf{U}_{p,k})= \mathbf{p}_{k}^{H}\mathbf{A}_{p,k}\mathbf{p}_{k}+ \sum_{i\neq k}^{K}\mathbf{p}_{i}^{H}\mathbf{A}_{p,k}\mathbf{p}_{i} - \mathbf{a}_{p,k}^{H}\mathbf{p}_{k}- \mathbf{p}_{k}^{H}\mathbf{a}_{p,k}+{\Phi}_{p,k},
\end{split}
\end{equation}
where $\mathbf{p}_{c} = vec(\mathbf{P}_{c})$, $\mathbf{p}_{k} = vec(\mathbf{P}_{k})$, $\mathbf{A}_{c,k}^{'} = \mathbf{I}_{Q_c}\otimes\mathbf{H}_{k}\mathbf{G}_{c,k}^H\mathbf{U}_{c,k}\mathbf{G}_{c,k}\mathbf{H}_{k}^{H}$, $\mathbf{A}_{z,k} = \mathbf{I}_{Q_{k}}\otimes\mathbf{H}_{k}\mathbf{G}_{z,k}^H\mathbf{U}_{z,k}\mathbf{G}_{z,k}\mathbf{H}_{k}^{H}$, $\mathbf{a}_{z,k} = vec(\mathbf{U}_{z,k}\mathbf{H}_{k}\mathbf{G}_{z,k}^H)$ and ${\Phi}_{z,k} = \sigma_{n}^{2}tr(\mathbf{U}_{z,k}\mathbf{G}_{z,k}\mathbf{G}_{z,k}^H) + tr(\mathbf{U}_{z,k})-\log\det(\mathbf{U}_{z,k})$, $z\in\{c,p\}$.
Next, we calculate the SAFs of the AWMSEs following (\ref{eq:a14}) and (\ref{eq:15}).
\subsubsection{STEP 1}
Let us denote the precoders from the previous iteration as $\mathbf{P}^{[i-1]}$. For each channel realization, the equalizers $\big(\mathbf{G}(\mathbf{P}^{[i-1]})\big)^{(n)}$ and weights $\big(\mathbf{U}(\mathbf{P}^{[i-1]})\big)^{(n)}$ are calculated for both common and private stream vectors. Precoders are fixed for each i.i.d realization. After obtaining equalizers and weights, we consider SAFs of the following entities: 
$\widehat{\mathbf{A}}_{c,k}^{'(N)}=\frac{1}{N}\sum_{n=1}^{N} (\mathbf{A}_{c,k}^{'})^{(n)}$, $\widehat{\mathbf{A}}_{z,k}^{(N)}=\frac{1}{N}\sum_{n=1}^{N} \mathbf{A}_{z,k}^{(n)}$, $\widehat{\mathbf{a}}_{z,k}^{(N)}=\frac{1}{N}\sum_{n=1}^{N} \mathbf{a}_{z,k}^{(n)}$, $\widehat{\Phi}_{z,k}^{(N)}=\frac{1}{N}\sum_{n=1}^{N} {\Phi}_{z,k}^{(n)}$, $z\in\{c,p\}$.
\subsubsection{STEP 2}
The next step is to update the precoders by substituting the updated equalizers, weights and SAFs of dependent entities into equation (\ref{eq:13}). The problem is formulated as 
\begin{subequations}\label{eq:16}
\begin{align}
&\min_{\mathbf{P},\widehat{\mathbf{x}}} \sum_{k=1}^{K} \mu_{k} \bigg(\widehat{X}_{k}+ \mathbf{p}_{k}^{H}\widehat{\mathbf{A}}_{p,k}\mathbf{p}_{k}+ \sum_{i\neq k}^{K}\mathbf{p}_{i}^{H}\widehat{\mathbf{A}}_{p,k}\mathbf{p}_{i} - \widehat{\mathbf{a}}_{p,k}^{H}\mathbf{p}_{k}- \mathbf{p}_{k}^{H}\widehat{\mathbf{a}}_{p,k}+\widehat{\Phi}_{p,k}\bigg) \\ 
&\sum_{i=1}^{K}\widehat{X}_{i} +Q_c \geq\mathbf{p}_{c}^{H}\widehat{\mathbf{A}}_{c,k}^{'}\mathbf{p}_{c}+ \sum_{i=1}^{K}\mathbf{p}_{i}^{H}\widehat{\mathbf{A}}_{c,k}\mathbf{p}_{i} - \widehat{\mathbf{a}}_{c,k}^{H}\mathbf{p}_{c}- \mathbf{p}_{c}^{H}\widehat{\mathbf{a}}_{c,k}+\widehat{\Phi}_{c,k},\,\,\forall\,k\,\in \mathcal{K}\\
&tr(\mathbf{P}\mathbf{P}^H) \leq {P_{t}}\\
&\widehat{\mathbf{x}}\leq \mathbf{0}.
\end{align}
\end{subequations}
\begin{algorithm}[b]
\caption{AO ALGORITHM}\label{alg:AO}
\begin{algorithmic}[1]
\State $\mathbf{Initialize}\:\: i=0,\: \mathbf{P}^{[0]}$
\State $\mathbf{Iterate}$
\State $\;\;\;i = i+1,\;\mathbf{G}=\mathbf{G}(\mathbf{P}^{[i-1]}),\:\mathbf{U}=\mathbf{U}(\mathbf{P}^{[i-1]}).$
\State $\;\;\;\mathrm{Compute}\;\widehat{\mathbf{A}}_{c,k}^{'},\, \widehat{\mathbf{A}}_{c,k},\, \widehat{\mathbf{A}}_{p,k},\,\widehat{\mathbf{a}}_{c,k},\,\widehat{\mathbf{a}}_{p,k},\,\widehat{\Phi}_{c,k},\,\widehat{\Phi}_{p,k},\;\;\;\;\forall\:k\in\:\mathcal{K}.$
\State $\;\;\;\mathrm{Solve}\,\,(\ref{eq:16}),\; \mathrm{update}\;\mathbf{P}^{[i]}, \widehat{\mathbf{x}}.$
\State \textbf{until} $convergence$
\end{algorithmic}
\end{algorithm}
\par Problem (\ref{eq:16}) is a convex Quadratically Constrained Quadratic Program (QCQP) and can be solved using interior-point methods \cite{ye1997interior,grant2008cvx}. Step 1 and 2 are repeated until the convergence is reached as specified in Algorithm 1. Proposition 1 of \cite{RS2016hamdi} and its proof shows that for a given $\mathbb{H}^{(N)}$, Algorithm 1 converges to a KKT solutions of the sampled WASR problem (\ref{eq:7}) and as $N\rightarrow \infty$, converges to a KKT solution of the WASR in problem (\ref{eq:3}).
\section{PHY-Layer Design for MIMO Channels}
\label{sec:transceiver}
\begin{figure}
\centering
\includegraphics[width=5.5in,height=3.2in]{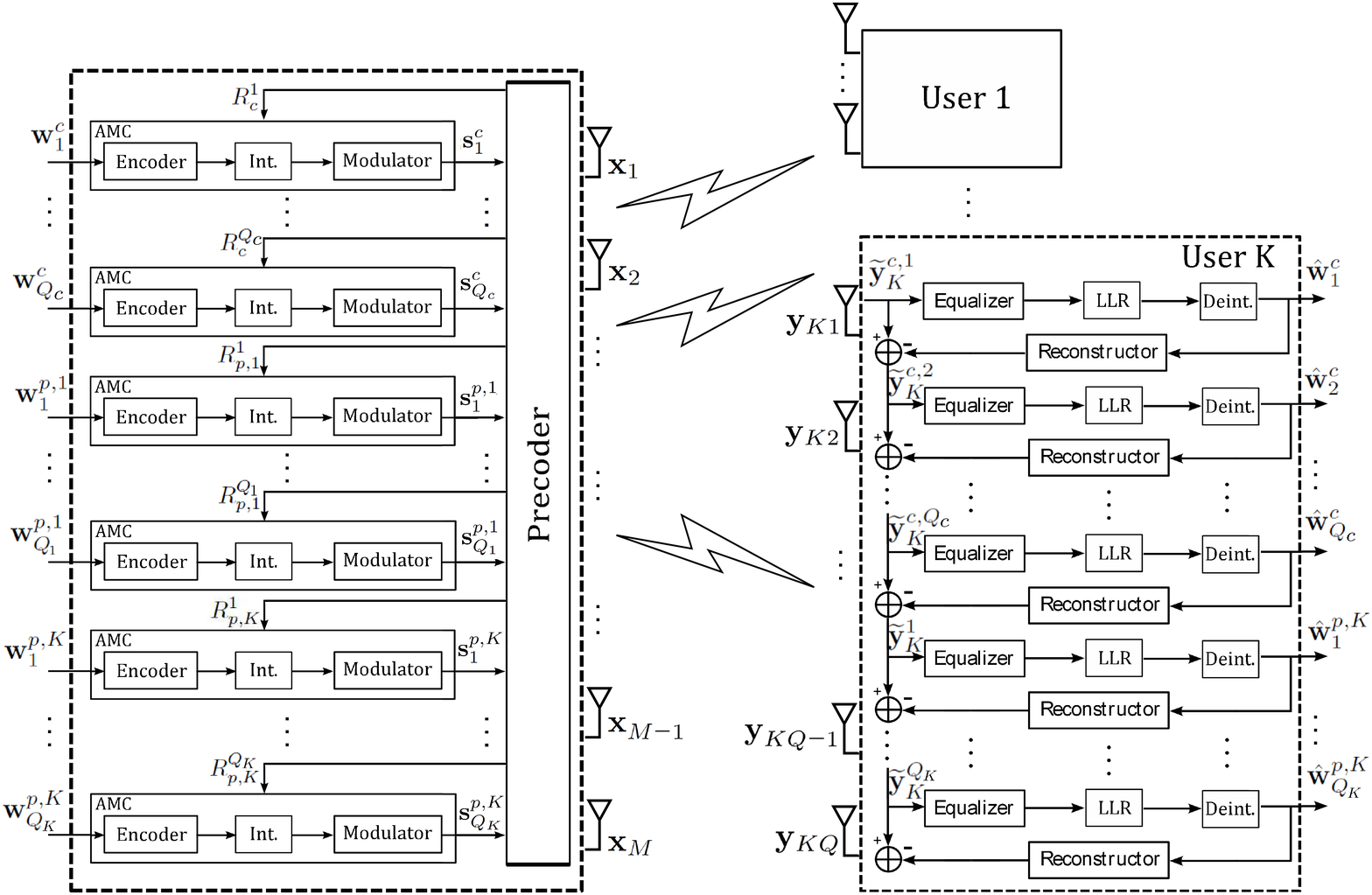}%
\caption{Proposed transceiver architecture}%
\label{fig:architecture}
\end{figure}
In addition to the theoretical foundations, it is of high importance to demonstrate the improved performance of RS in practical setups. In this section, we propose a practical transceiver architecture for RS in MIMO settings.
Fig.~\ref{fig:architecture} illustrates the proposed transceiver architecture, which is build upon and generalizes the design in \cite{dizdar_2020} of RS in MISO channels. 
The transmitter employs finite alphabet modulation schemes $4$-QAM, $16$-QAM, $64$-QAM and $256$-QAM, finite-length polar coding \cite{arikan_2009} for Forward Error Correction (FEC), and an Adaptive Modulation and Coding (AMC) algorithm. 

The combined common messages are mapped to binary vectors $\mathbf{w}^{c}_{i}$ of length $K^{c}_{i}$, for $i \in \left\lbrace 1, 2, \ldots, Q_{c}\right\rbrace$, respectively. Similarly, the private messages are respectively mapped to binary vectors $\mathbf{w}^{p,k}_{j}$ of length $K^{p,k}_{j}$, for $k \in \mathcal{K}$ and $j \in \left\lbrace 1, 2, \ldots, Q_{k}\right\rbrace$.
We assume the split messages are independent, such that, the common and private information bit vectors are independent and uniformly distributed in $\mathbb{F}_{2}^{K^{c}_{i}}$ and $\mathbb{F}_{2}^{K^{p,k}_{j}}$.
The information bit vectors are independently encoded and modulated into common and private symbol streams $\mathbf{s}^{c}_{i}$ and $\mathbf{s}^{p,k}_{j}$, each of length $S$. 
The AMC algorithm selects a suitable modulation-coding rate pair based on the ARs. In this work, the transmit rate calculations for the AMC algorithm are performed assuming the instantaneous channel is known at the AMC module. More details on the channel coding procedure and the AMC algorithm are given in \cite{dizdar_2020} for the interested reader. The precoders for the common and private streams are obtained as described in Algorithm~\ref{alg:AO}. 

We note that the rate expressions in (\ref{eq:InstRate_PC}) are valid under the assumption of joint decoding of all $Q_c$ common streams (and all $Q_{k}$ private streams, $\forall k \in \mathcal{K}$). This restricts the use of conventional and practical point-to-point decoding methods for channel coding at the receiver side. Although there are studies on joint decoding of several types of channel codes ({\sl} e.g., polar codes, Low-Density Parity Check codes), such implementations have higher complexities than point-to-point decoding methods, especially when the number of jointly decoded streams increase. 

Instead of performing joint decoding, we perform interference nulling and interference cancellation among the streams at the receiver in order to benefit from low-complexity decoding methods. Such receiver design is originally proposed in \cite{wolniansky_1998, foschini_1999} for V-BLAST systems. The proposed design allows to obtain a separate transmission rate for each common and private stream for Modulation and Coding Scheme (MCS) selection, as opposed to assigning a single rate value calculated by (\ref{eq:InstRate_PC}) to all common and private streams.  
The proposed receiver architecture is illustrated in Fig.~\ref{fig:architecture} for a detection and decoding ordering based on the natural ordering of the stream indexes. We note that the detection and decoding ordering of the common (and private) streams in the figure and the following explanations is for the sake of simplicity and the actual stream ordering criterion we use in our design is also explained in this section. 

Consider the scenario where the common streams $1,2,\ldots,l-1$, for any $l < Q_c$, have been correctly decoded and removed from the received signal at user-$k$ to obtain the resulting interference cancelled received signal $\widetilde{\mathbf{y}}_{k}^{l}$.
We define the effective channel for the $l$-th common stream as \mbox{$\bar{\mathbf{h}}_{c,k}^{(l)}\triangleq\mathbf{H}_{k}^{H}\mathbf{P}_{c}(l) \in \mathbb{C}^{Q\times1}$}, where $\mathbf{P}_{c}(l)$ is the $l$-th column of the matrix $\mathbf{P}_{c}$.
We can write $\widetilde{\mathbf{y}}_{k}^{l}$ in terms of the real and effective channels as 
\begin{equation}
\widetilde{\mathbf{y}}_{k}^{c,l}=\sum_{i=l}^{Q_c}\bar{\mathbf{h}}_{c,k}^{(i)}s_{i}^{c}+\sum_{j\in\mathcal{K}}\mathbf{H}_k^{H}\mathbf{P}_{j}\mathbf{s}_{j}+\mathbf{n}_k.
\end{equation}
The detection of the $l$-th common stream is performed by multiplying $\widetilde{\mathbf{y}}_{k}^{c,l}$ with a linear nulling filter, $\mathbf{g}_{c,k}^{l}$. 
The nulling filter is designed based on the MMSE criterion and expressed as
\begin{align}
    \mathbf{g}_{c,k}^{l}=(\bar{\mathbf{h}}_{c,k}^{(l)})^{H}\hspace{-0.1cm}\left(\mathbf{I}_{Q}\hspace{-0.1cm}+\hspace{-0.1cm}\sum_{i=l}^{Q_c}\bar{\mathbf{h}}_{k}^{(l)}(\bar{\mathbf{h}}_{k}^{(l)})^{H}\hspace{-0.1cm}+\hspace{-0.1cm}\sum_{j\in\mathcal{K}}\mathbf{H}_{k}^{H}\mathbf{P}_{j}\mathbf{P}_{j}^{H}\mathbf{H}_{k}\right)^{-1}\hspace{-0.4cm}. \nonumber
\end{align} 
The definitions above use the assumption that the detection order follows the natural indexing of the streams for the brevity of the explanations. It is demonstrated in \cite{wolniansky_1998, foschini_1999} that ordering the streams according to their post-processing SINR yields the best performance. Therefore, we follow such approach by calculating the post-processing SINRs over the streams which are filtered by their corresponding linear nulling filters.
Consider the detection and decoding of the $l$-the common stream at user-$k$. The index $i^{\prime}$ of such stream is determined as the solution of the problem
\begin{align}
    i^{\prime}=\argmax_{i \in \mathcal{S}_{l}}\gamma^{i}_{c,k}, \nonumber
\end{align}
where $\mathcal{S}_{l}$ is the index set of the undetected streams with a cardinality of $|\mathcal{S}_{l}|=Q_{c}-l+1$, \mbox{$\gamma^{i}_{c,k}=1/\epsilon^{i}_{c,k}-1$} is the post-processing SINR of the undetected common stream $i$ at user-$k$ and  $\epsilon^{i}_{c,k}\triangleq \mathbb{E}\left\lbrace |\mathbf{g}_{c,k}^{i}\widetilde{\mathbf{y}}_{k}^{c,i}-s_{c,i}|^{2}\right\rbrace$ is the MSE of the undetected common stream $i$.

The decoding of the common stream $l$ is performed by a Soft Decision (SD) decoder, for which the Log-Likelihood Ratios (LLRs) are calculated over the equalized symbol $\mathbf{g}_{c,k}^{l}\widetilde{\mathbf{y}}_{k}^{c,l}$.
We use the LLR calculation method in \cite{seethaler_2004}. Let $\lambda_{c,k,i}^{l}$ denote the LLR of the $i$-th bit of the equalized common stream symbol $l$ for the $k$-th user. We write
\begin{align}
\lambda_{k,i}^{c,l}=\gamma_{c,k}^{l}\left[\min_{a\in \theta_{1}^{(i)}}\psi(a)-\hspace{-0.2cm}\min_{a\in \theta_{0}^{(i)}}\psi(a) \right], \nonumber
\end{align}
\hspace*{-0.05cm}where $\theta_{b}^{(i)}$ is the set of modulation symbols with the value $b$, $b \in \{0,1\}$ at the $i$-th bit location, \mbox{$\psi(a)=|\frac{\mathbf{g}_{c,k}^{l}\widetilde{\mathbf{y}}_{k}^{c,l}}{\rho^{l}_{c,k}}\hspace{-0.1cm}-a|^{2}$} and \mbox{$\rho^{l}_{c,k}=\gamma^{l}_{c,k}/(1+\gamma^{l}_{c,k})$}.

A similar procedure is applied to the private streams intended to user-$k$ after all common streams are decoded. The decoding operation is performed using a Successive Cancellation List (SCL) decoder for point-to-point polar codes \cite{tal_2015}.
\par \textit{Remark:} As our aim is to verify the theoretical foundations in the paper, we propose a receiver architecture which has a higher complexity than a receiver with a single interference cancellation (IC) process. An example design for a receiver with single IC would employ linear equalizers to detect each common (or private) stream by treating all other streams as interference and then decode all common (or private) streams in parallel. Although such receiver design is expected to suffer a performance loss due to its sub-optimality, it may be more suitable for practical systems due to its reduced complexity.
\section{Numerical Results}\label{NumRes}
In this section, we evaluate the performance of the RS in MIMO BC with perfect and imperfect CSIT. In the following, we first illustrate the WSR performance of RS in MIMO BC followed by the sum-DoF performance. At last, we illustrate the LLS results. For comparison, the following three strategies are considered as baselines:
\begin{itemize}
\item{\textbf{DPC}}: Implemented based on the algorithm in \cite{DPCregion}. With perfect CSIT, DPC is a capacity achieving scheme which cancels interference at the transmitter.
\item{\textbf{MU--MIMO}}: Results are produced by turning off the common stream and solving problem (\ref{eq:3}), i.e, allocating no power to the common stream vector.
\item{\textbf{MIMO NOMA}}: Implemented by extending the degraded beamforming methodology proposed in \cite{hamdi2017bruno} to the MIMO case, delineated in Appendix A. The precoders and the decoding order are jointly optimized to achieve maximum performance, where precoders are optimized using the WMMSE and AO algorithm.
\end{itemize}
Note that the comparison with DPC is limited to the perfect CSIT scenario (as DPC is capacity achieving only with perfect CSIT) and MIMO NOMA is limited to the $2$-user case because of the high complexity in joint optimization of decoding order and precoders for more than $2$ users. User channels are randomly generated in accordance with \cite{RS2016hamdi,hamdi2016robust}. The actual channel experienced at user-$k$ $\mathbf{H}_k$ and the channel estimation error  $\widetilde{\mathbf{H}}_{k}$ both have complex Gaussian entries drawn from the distribution  $\mathcal{C}\mathcal{N}({0},\sigma_{k}^{2})$ and $\mathcal{C}\mathcal{N}({0},\sigma_{e,k}^{2})$ respectively. The channel estimation error power is defined as $\sigma_{e,k}^{2} \triangleq \sigma_{k}^{2}P_{t}^{-\alpha}$ such that the CSIT quality for user channels scale with both channel variance and transmit power. Consequently, the channel estimate $\widehat{\mathbf{H}}_{k} = \mathbf{H}_k-\widetilde{\mathbf{H}}_{k}$ also follows Gaussian distribution with entries $\mathcal{C}\mathcal{N}({0},\sigma_{k}^{2}-\sigma_{e,k}^{2})$. By averaging the WASR over $100$ channel realizations, WESR is obtained. For each channel estimate $\widehat{\mathbf{H}}$, $N=1000$ channel error samples are generated to form $\mathbb{H}^{(n)}$  and then the SAA method is used to approximate the AR. For a given channel estimate $\widehat{\mathbf{H}}$, the $n^{th}$ channel estimation error $\widetilde{\mathbf{H}}^{(n)}$ sample is generated randomly from the error distribution and forms the $n^{th}$ conditional channel ${\mathbf{H}}^{(n)}=\widehat{\mathbf{H}}^{(n)}+\widetilde{\mathbf{H}}^{(n)}$. In the case of perfect CSIT, $N=0$ and $\widehat{\mathbf{H}}=\mathbf{H}$.
\par Initialization of precoders of all  three schemes is crucial and plays an important role at higher SNRs, especially for convergence\cite{RS2016hamdi}. For MU--MIMO and MIMO NOMA, the precoders are initialized using Maximum Ratio Transmission (MRT). Initial power allocation is uniform among the users. For RS, the initialization of precoders is according to MRT-Singular Value Decomposition (SVD) in which the private streams are initialized using MRT and the initial precoder for the common stream vector is the dominant $M\times Q_c$ sub-matrix of the left singular matrix of $\widehat{\mathbf{H}}$. Power distribution is $q_c=P_t -P_t^{\alpha}$ for the common stream vector and the remaining power $P_t^{\alpha}$ is uniformly distributed among the private stream vectors of all users. The noise variance is assumed to be $\sigma_{n}^{2}=1$.
\subsection{WESR Performance: Rate-Region}
\begin{figure}
\centering
\subfloat[$\sigma_{2}^{2}=1$]{\includegraphics[width=5.5cm,height=4.5cm]{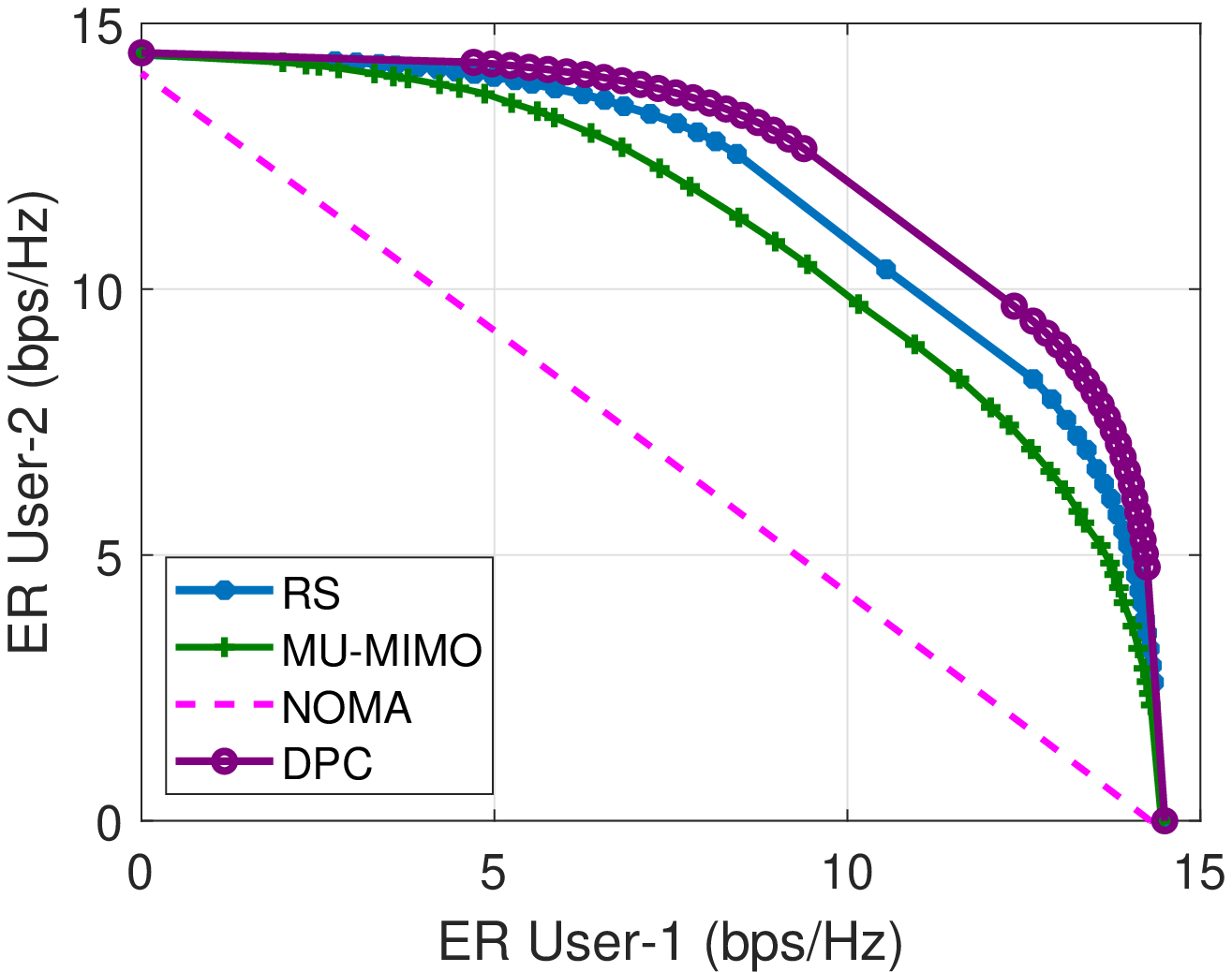}}%
\subfloat[$\sigma_{2}^{2}=0.09$]{\includegraphics[width=5.5cm,height=4.5cm]{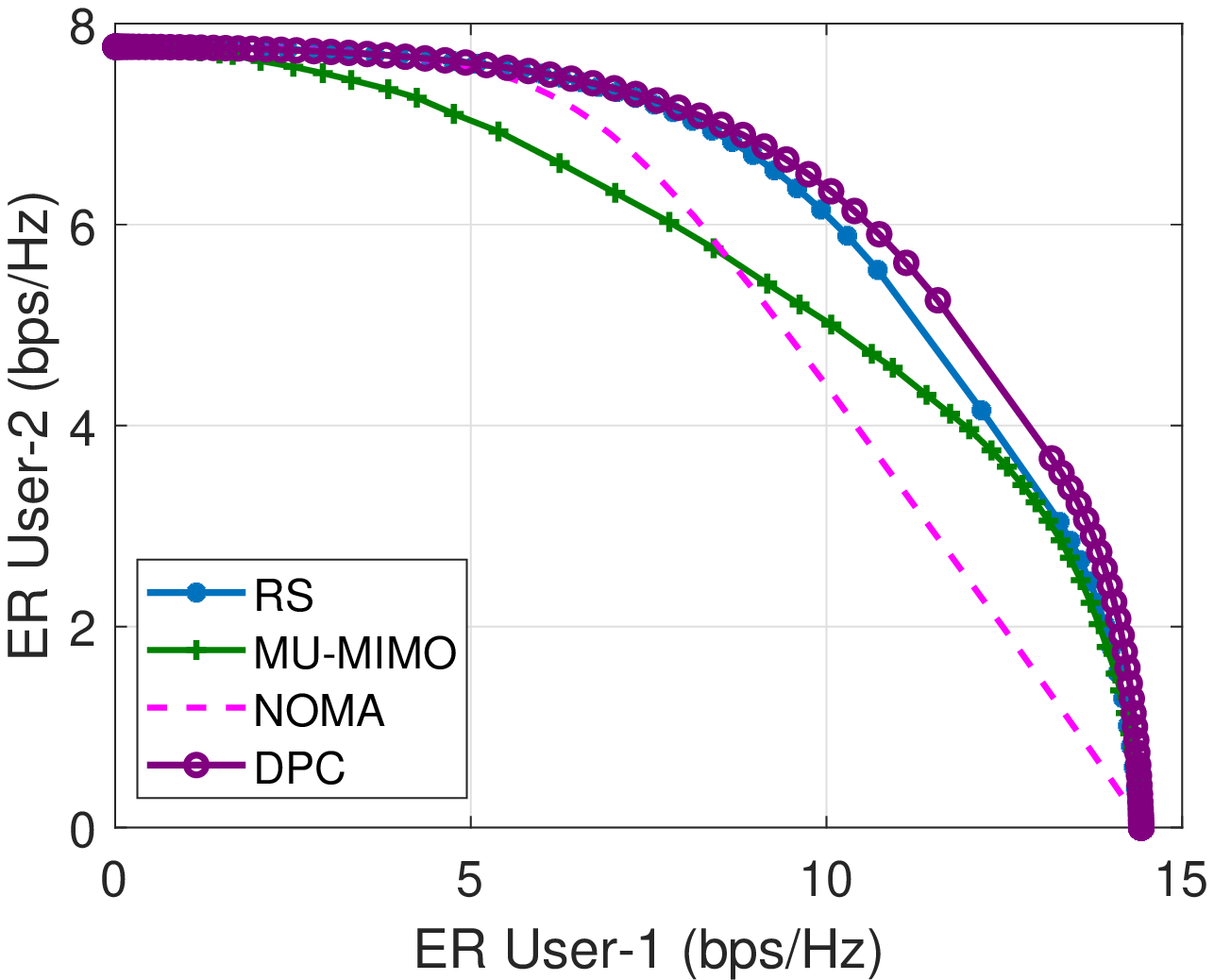}}
\caption{ER-region comparison of different strategies with perfect CSIT, averaged over $100$ random channel realizations, SNR$=20$ dB, $M=4$, $K=2$, $Q=2$, $Q_{c}=2$, $\sigma_{1}^{2}=1$.}
\label{fig:ER_PerfectCSIT}
\end{figure}

\begin{figure}
\centering
\subfloat[$\sigma_{2}^{2}=1$]{\includegraphics[width=5.5cm,height=4.5cm]{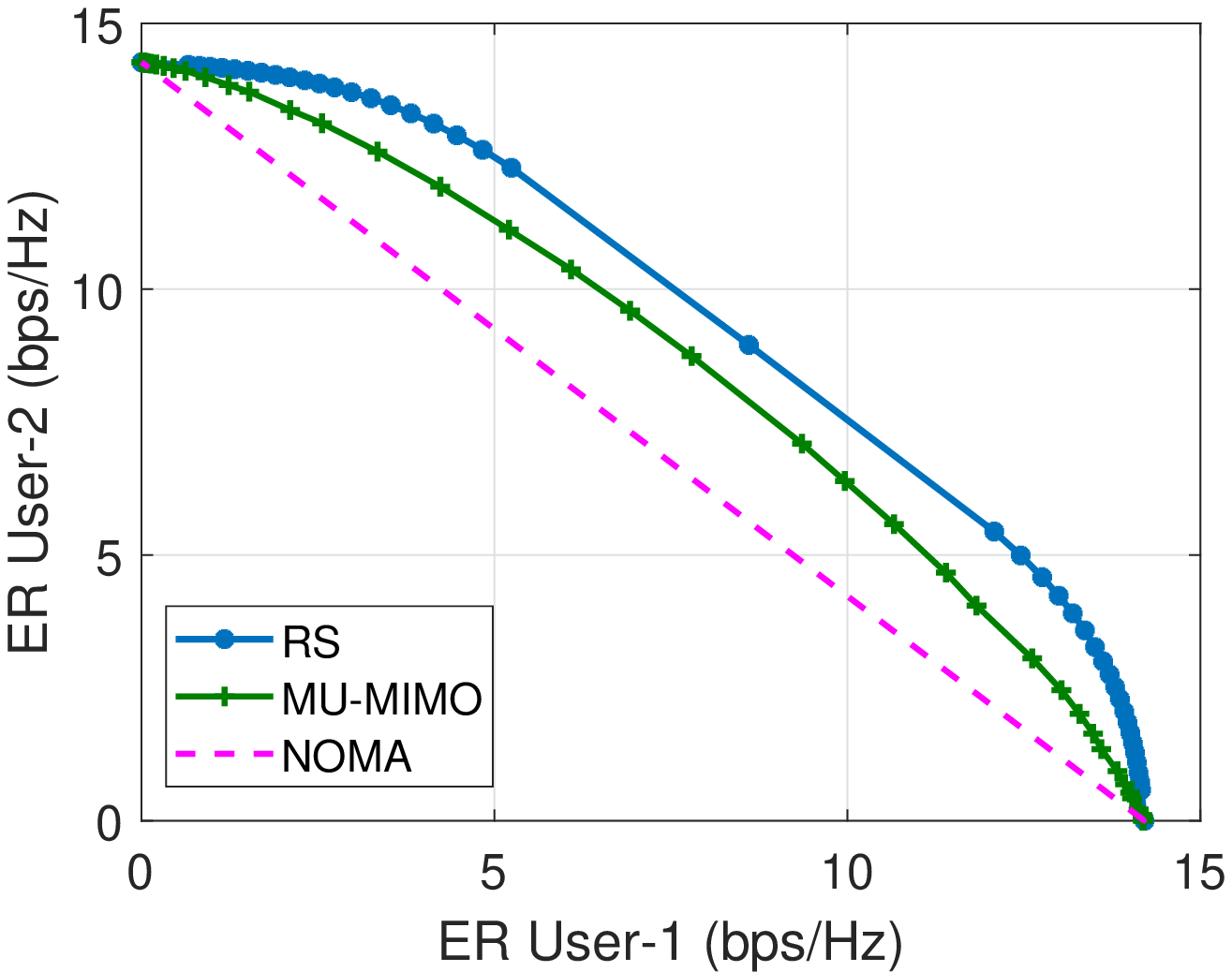}}%
\subfloat[$\sigma_{2}^{2}=0.09$]{\includegraphics[width=5.5cm,height=4.5cm]{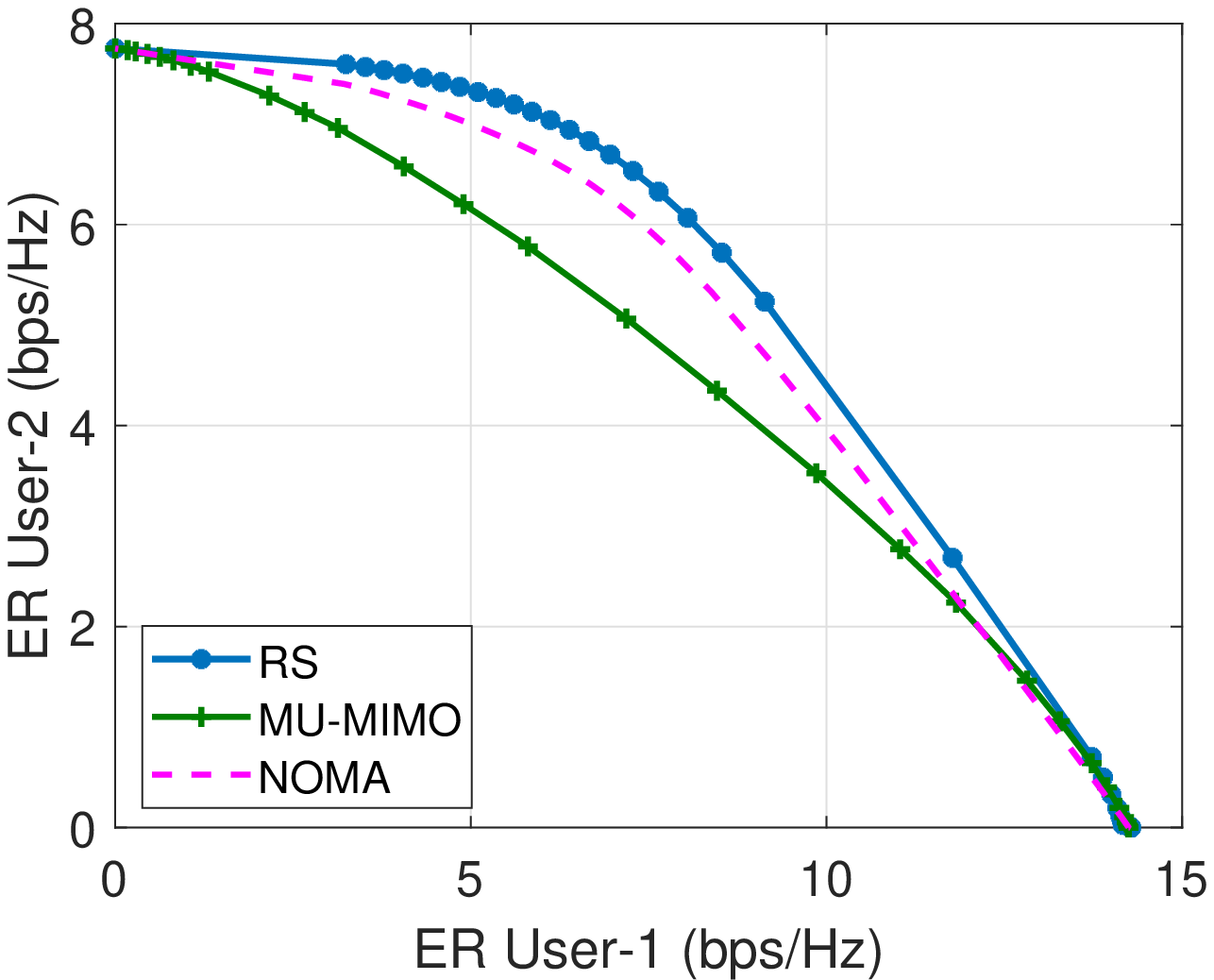}}
\caption{ER-region comparison of different strategies with imperfect CSIT, averaged over $100$ random channel realizations, SNR$=20$ dB, $M=4$, $K=2$, $Q=2$, $Q_{c}=2$, $\alpha=0.6$, $\sigma_{1}^{2}=1$.}
\label{fig:ER_PartialCSIT}
\end{figure}

We first consider $M=4$, $K=2$, $Q=2$, $Q_c=2$ and SNR=$20$ dB. The ER-regions achieved by different strategies are illustrated for perfect and imperfect CSIT in Fig.~\ref{fig:ER_PerfectCSIT} and Fig.~\ref{fig:ER_PartialCSIT}, respectively, where different channel strength disparities are considered for analysis. A boundary point for any transmission strategy is realized by solving the WESR problem for a weight pair by averaging the WASR over $100$ channel realizations such that for each channel realization, we use Algorithm $1$ to obtain the WASR for that strategy. The entire rate region is calculated over a set of different weight pairs assigned to users. To obtain the rate-regions, the weight of user-$1$ is fixed at $\mu_{1}=1$ and the weight of user-$2$ is varied as $\mu_2\in10^{[-3,-1,-0.95,\ldots,0.95,1,3]}$. 
\par Fig.~\ref{fig:ER_PerfectCSIT} illustrates the ER-region of all the four transmission strategies in the perfect CSIT scenario. In both subfigures (a) and (b), DPC achieves the highest rate-region, which is the capacity region. In Fig.~\ref{fig:ER_PerfectCSIT}(a), we observe that with no disparity in the strength of user channels, MIMO NOMA achieves the worst rate region. As the MIMO NOMA strategy is motivated to exploit disparities in channel strengths, it is unable to properly manage the interference in this scenario. Whereas, MU--MIMO achieves a larger rate region compared to MIMO NOMA as it depends on the precoder design at the transmitter. At the receiver side, each user decodes its own streams by treating the streams of other users as noise. In contrast, when user-$2$ suffers an additional $10$ dB path loss, MIMO NOMA achieves a larger rate-region as illustrated in Fig.~\ref{fig:ER_PerfectCSIT}(b) compared to MU--MIMO when the  weight of user-$2$ is either more than or comparable to the weight of user-$1$. When the weight of user-$1$ is significantly larger than the weight of user-$2$, the effect of disparities in channel strength fades as user-$2$ is weighted significantly less. Hence, MU--MIMO starts performing better than MIMO NOMA. In comparison, RS performs better than both MU--MIMO and MIMO NOMA and achieves a closer rate region to the capacity achieving DPC in both subfigures. This is due to the fact that it allows the users to exploit the common streams thereby enabling them to partially decode the interference and partially treat the interference as noise.   
\par Fig.~\ref{fig:ER_PartialCSIT} illustrates the ER-regions obtained for the imperfect CSIT scenario. As the CSIT quality degrades, the performance of MU-MIMO deteriorates significantly while RS, because of the presence of common streams, exhibits robustness and shows explicit performance gain over MU-MIMO and MIMO NOMA. In the case of different channel strengths, we observe that the results are consistent with the perfect CSIT scenario. Comparing Fig.~\ref{fig:ER_PerfectCSIT} and Fig.~\ref{fig:ER_PartialCSIT}, we obtain that better management of interference makes RS robust to CSIT inaccuracy and different user deployments. 
\subsection{Sum-DoF: Effect of Common Message}
\subsubsection{CSIT Quality}
\begin{figure}
\centering
\subfloat[$\alpha=0.9$]{\includegraphics[width=5.2cm,height=4.2cm]{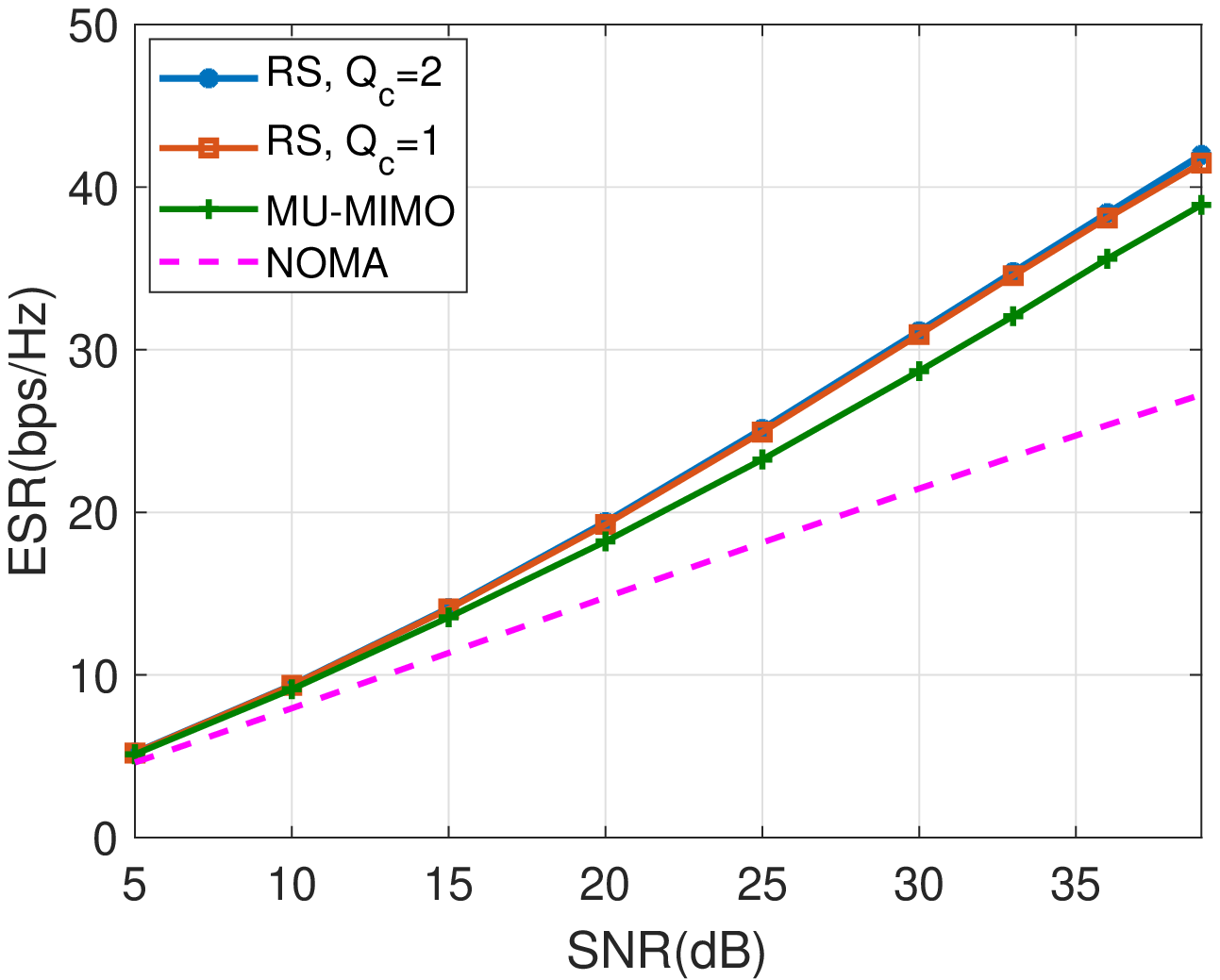}}
\subfloat[$\alpha=0.6$]{\includegraphics[width=5.2cm,height=4.2cm]{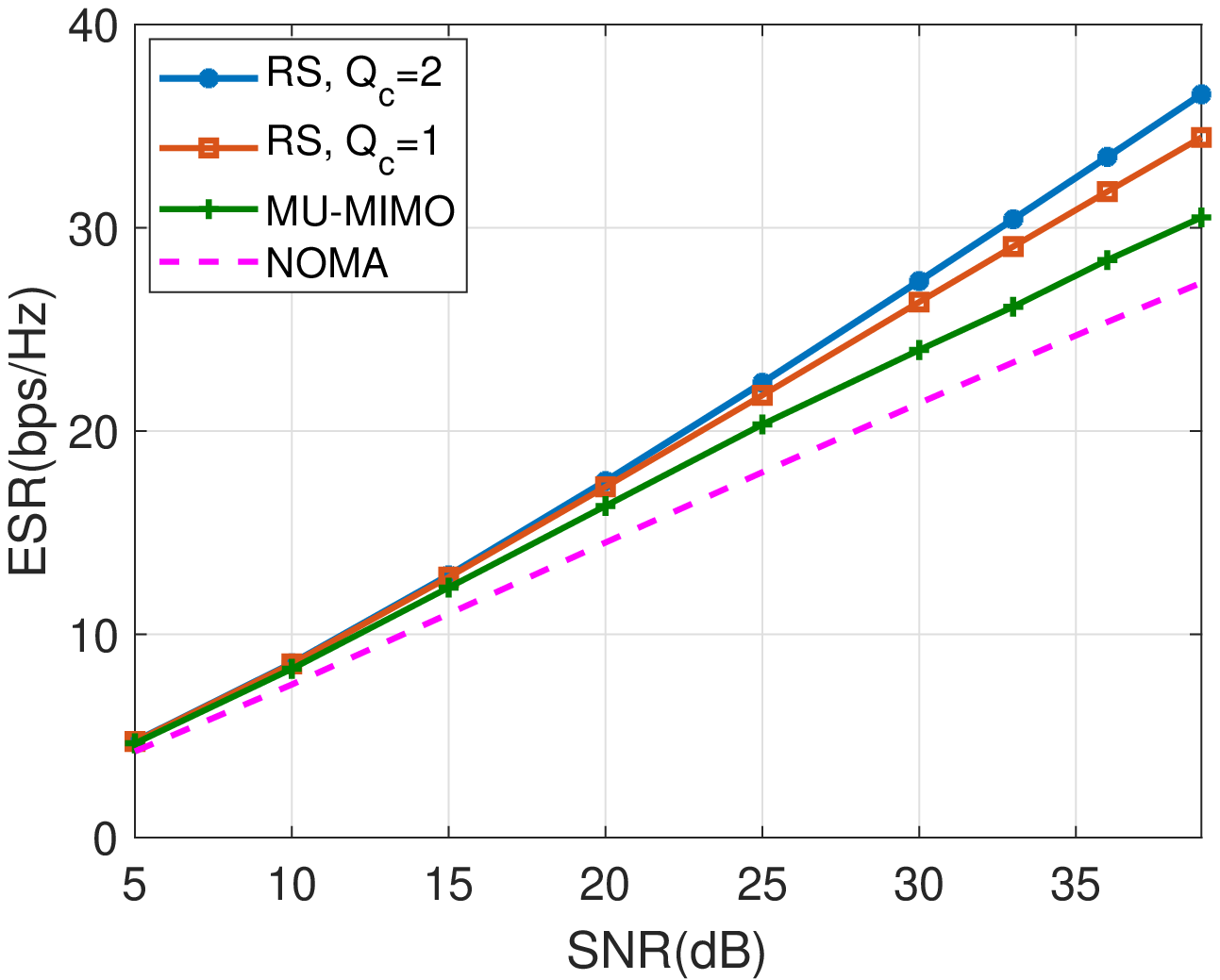}}
\subfloat[$\alpha=0.3$]{\includegraphics[width=5.2cm,height=4.2cm]{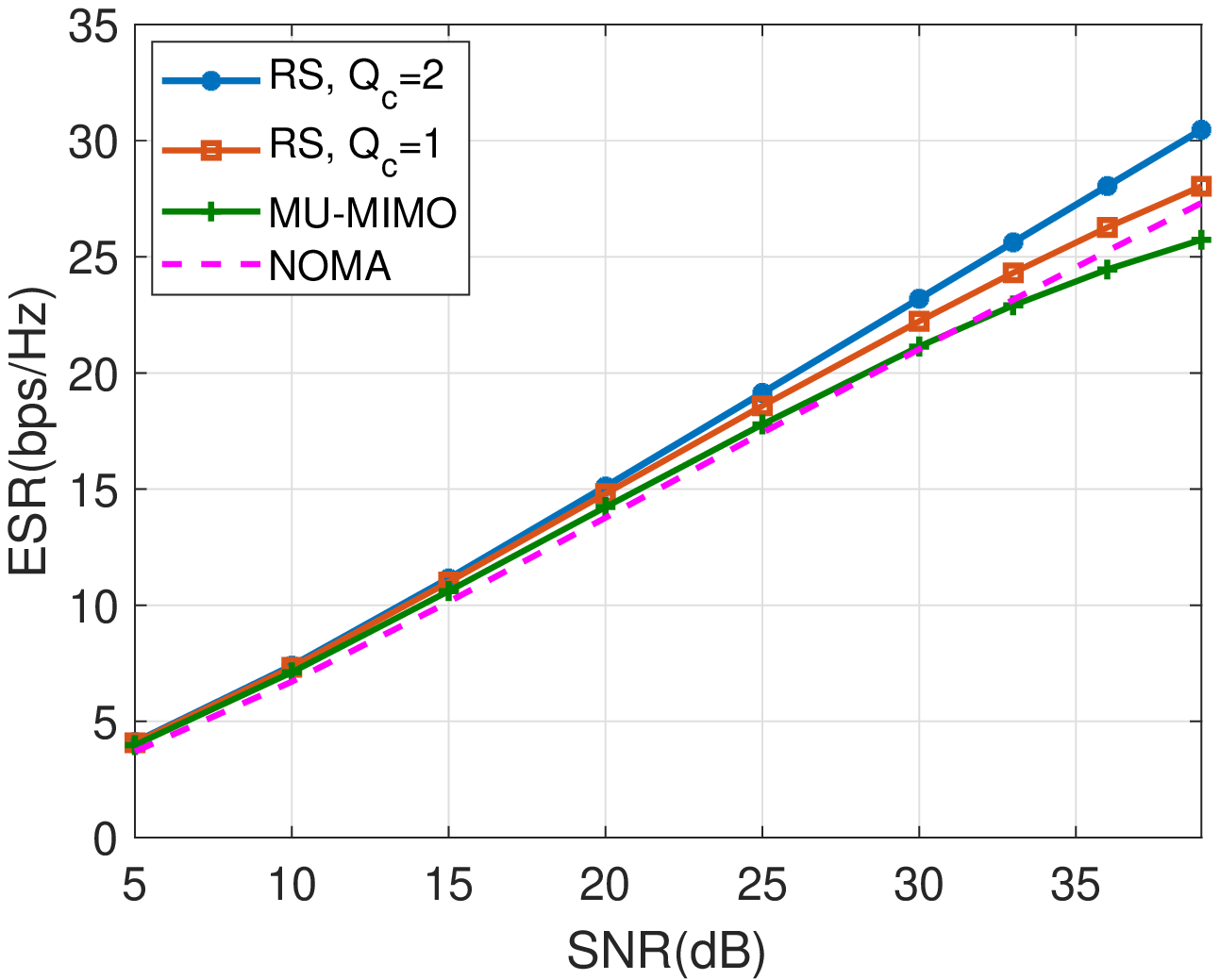}}
\caption{ESR versus SNR comparison of different strategies with different imperfect CSIT inaccuracies, averaged over 100 random channel realizations, $M=4$, $K=2$, $Q=2$, $\sigma_{1}^{2}=1$, $\sigma_{2}^{2}=1$.}
\label{fig:Alpha_PartialCSIT}
\end{figure}

From Fig.~\ref{fig:Alpha_PartialCSIT}(a), we see that at high SNR, RS with $Q_c=2$ always achieves a higher ESR performance compared to RS with $Q_c=1$, and both perform better than MU--MIMO and MIMO NOMA\footnote{Note that, the performance gain of RS over MIMO NOMA with $Q_{c}=1$ has implications on the receiver complexity with RS requiring less number of SICs compared to MIMO NOMA.}. For $\alpha=0.6$, the sum-DoF obtained for RS with $Q_{c}=2$, RS with $Q_{c}=1$, MU-MIMO and MIMO NOMA are $[3.08, 2.67, 2.10, 1.93]$, respectively. These values are close to the theoretical sum-DoF values $[3.2, 2.8, 2.4, 2]$ calculated using (\ref{eq:4})--(\ref{eq:nomadof}). Thus, transmitting multiple common streams yields better ESR and sum-DoF performance than transmitting a single common stream. Furthermore, as $\alpha$ decreases from $0.9$ to $0.3$, the ESR performance and the sum-DoF gain gaps between RS and the other two schemes increase, as seen in Fig.~\ref{fig:Alpha_PartialCSIT}(b) and Fig.~\ref{fig:Alpha_PartialCSIT}(c). The results are inline with equations (\ref{eq:4})--(\ref{eq:nomadof}). Since $\alpha=0.9$ is closer to the perfect CSIT case, at higher SNRs, the MU--MIMO curve is nearly parallel to the RS curve though with lower ESR. It is because the contribution of the common streams decreases with increase in $\alpha$ and w.r.t sum-DoF, MU--MIMO gets closer to RS. However, optimization and contribution of common stream still make the ESR performance of RS better than MU--MIMO. MIMO NOMA on the other hand is observed to have the same sum-DoF irrespective of $\alpha$. The observation is consistent with the theoretical sum-DoF expression in (\ref{eq:nomadof}) which clearly implies that the sum-DoF of MIMO NOMA is independent of $\alpha$ in our scenario\footnote{The results are consistent with the findings in \cite{clerckx2021noma} that NOMA is unable to exploit the available CSIT efficiently.}. Therefore, RS is more suited in MU multi-antenna networks than MU--MIMO and MIMO NOMA in MIMO BC, especially with the deteriorating CSIT quality.
\subsubsection{Number of Users}
\begin{figure}
\centering
\subfloat[$Q=2$, $M=KQ$ ]{\includegraphics[width=5.2cm,height=4.2cm]{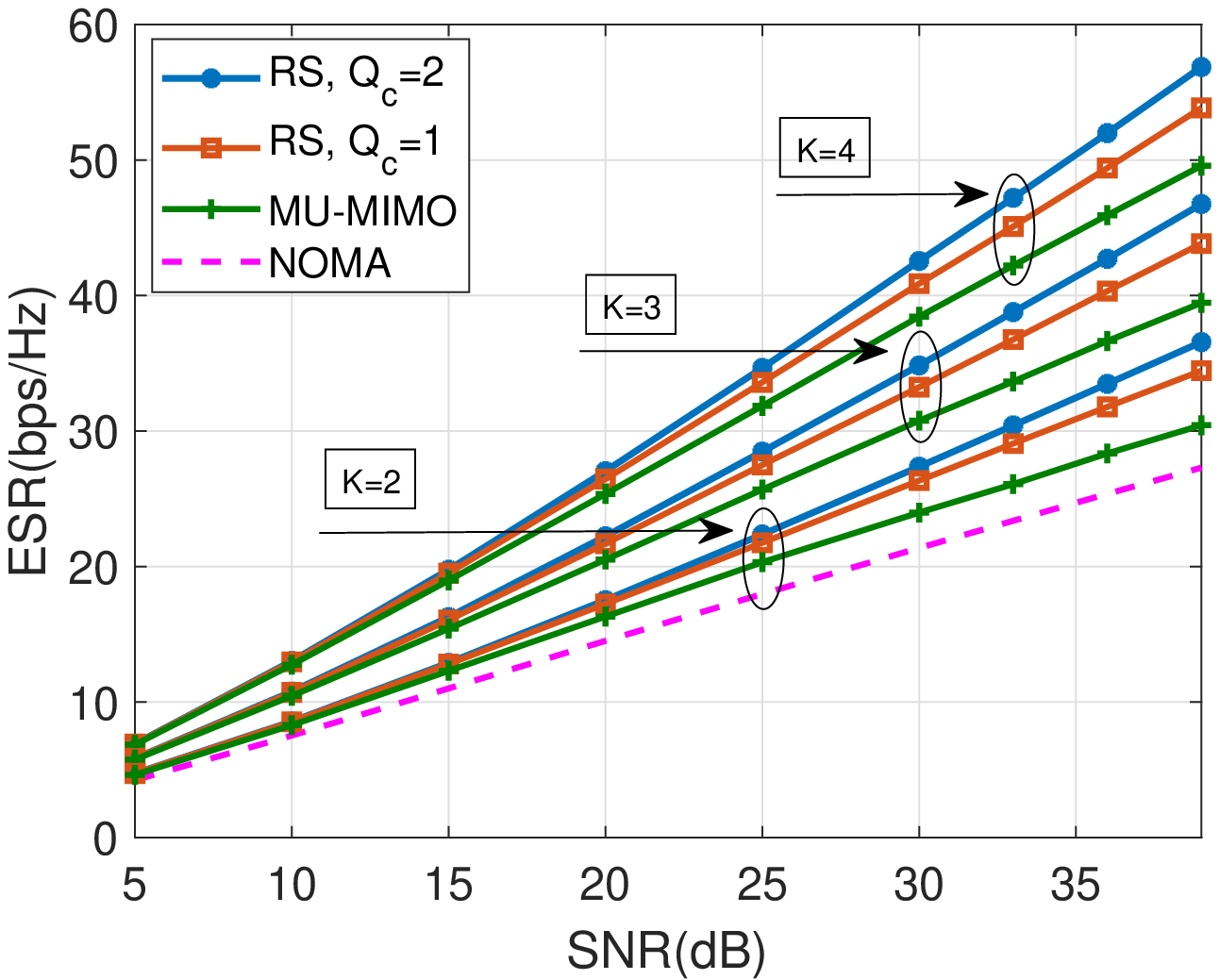}}
\subfloat[$Q=2$, $M=(K-1)Q$]{\includegraphics[width=5.2cm,height=4.2cm]{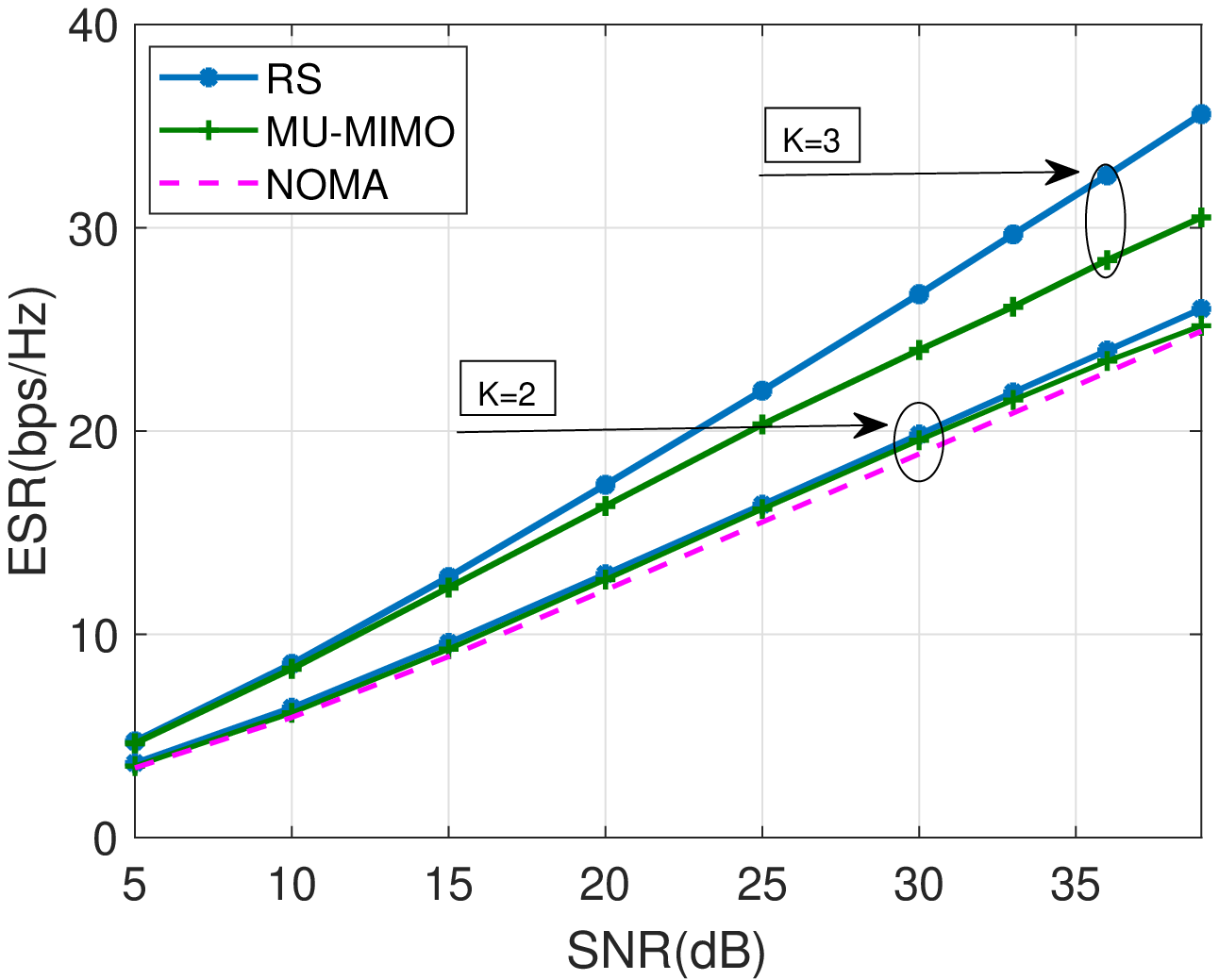}}
\subfloat[$Q=3$, $M=9$]{\includegraphics[width=5.2cm,height=4.2cm]{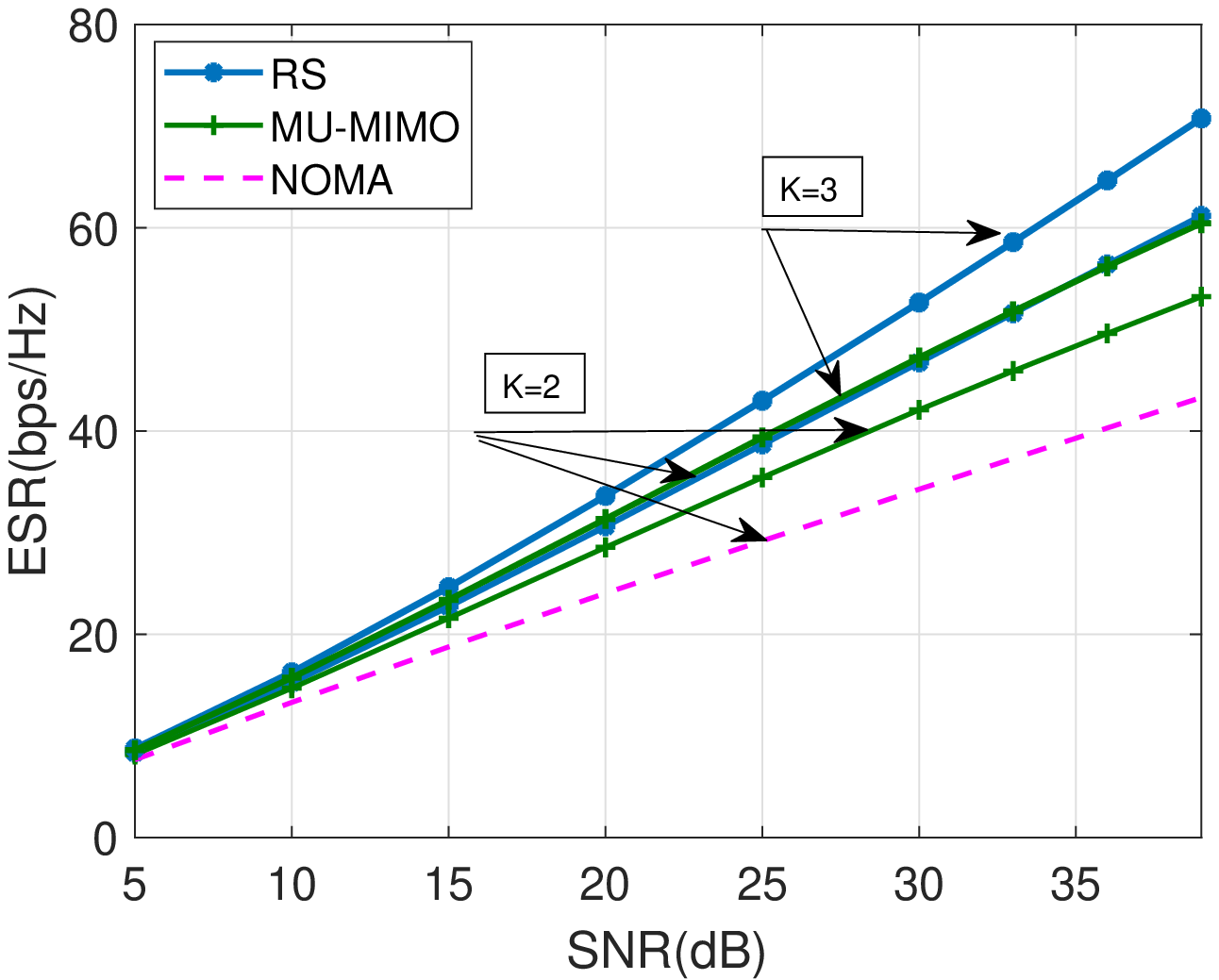}}
\caption{ESR versus SNR comparison of different strategies for different network loads, averaged over 100 random channel realizations, $\alpha=0.6$, $\sigma_{k}^{2}=1, \forall k\in[1,4]$.}%
\label{fig:UnderVsOverloaded_PartialCSIT}
\end{figure}

Fig.~\ref{fig:UnderVsOverloaded_PartialCSIT}(a) and Fig.~\ref{fig:UnderVsOverloaded_PartialCSIT}(b) illustrate the ESR performances of the schemes in the underloaded and overloaded regime, respectively. In the underloaded regime, for all the three user configurations, i.e., $K=2$, $K=3$ and $K=4$, RS with $Q_c=2$ has better ESR than RS with $Q_c=1$ which in turn performs better than MU--MIMO and MIMO NOMA. Fig.~ \ref{fig:UnderVsOverloaded_PartialCSIT}(b) illustrates that in the overloaded regime also, RS outperforms MU--MIMO and MIMO NOMA in ESR and sum-DoF performances. The sum-DoF achieved by all the three schemes are inline with equations (\ref{eq:4})--(\ref{eq:nomadof}) for the underloaded and overloaded scenarios.  Fig.~ \ref{fig:UnderVsOverloaded_PartialCSIT}(c) illustrates the ESR performance of all the three schemes with higher dimensions, i.e., $M=9,\, Q=3$. RS achieves ESR performance gain over MU--MIMO and MIMO NOMA in all scenarios.
\subsection{Link-Level Simulation Results}
\begin{figure}
\centering
\subfloat[$K=2$]{\includegraphics[width=5.5cm,height=4.5cm]{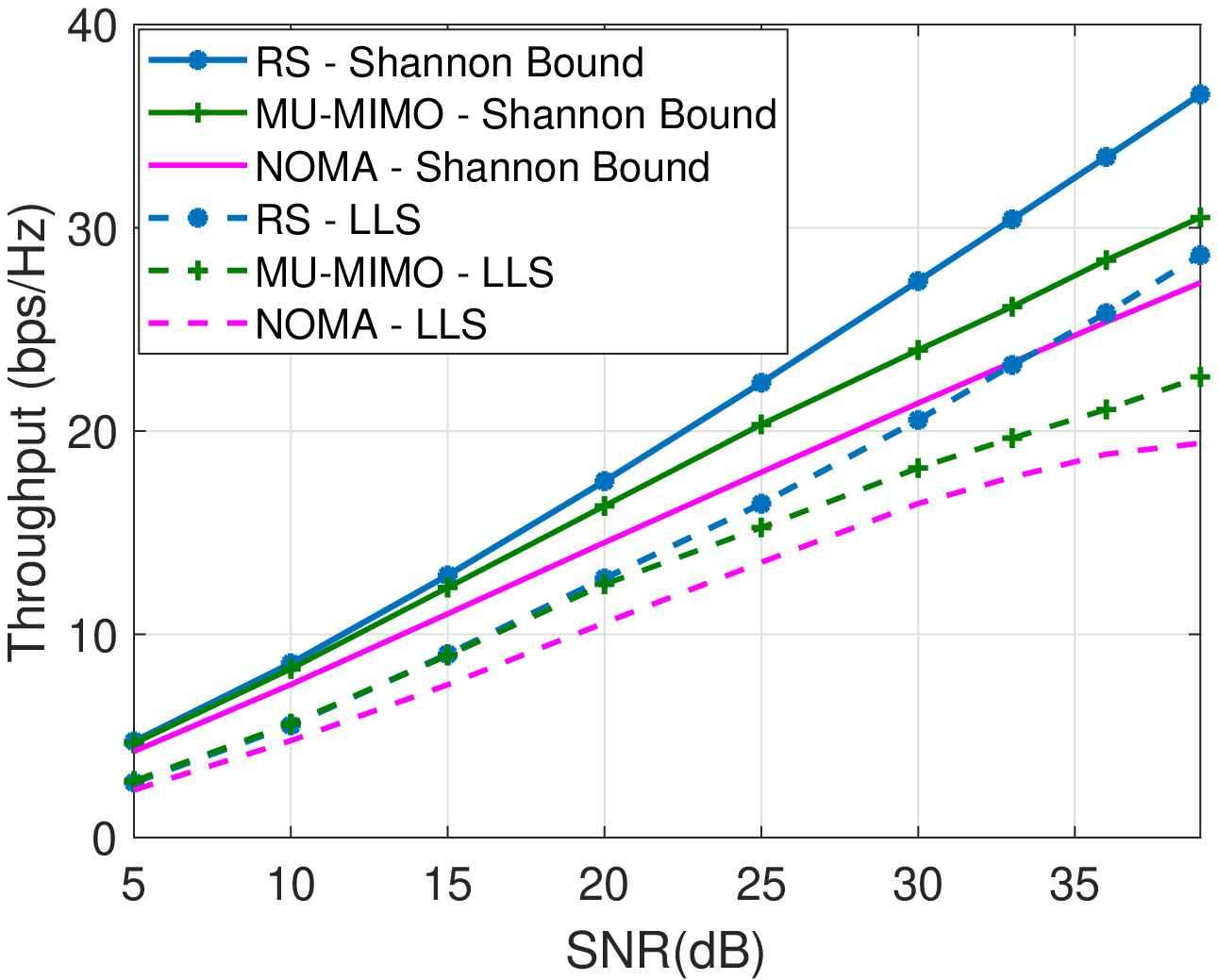}}%
\subfloat[$K=3$]{\includegraphics[width=5.5cm,height=4.5cm]{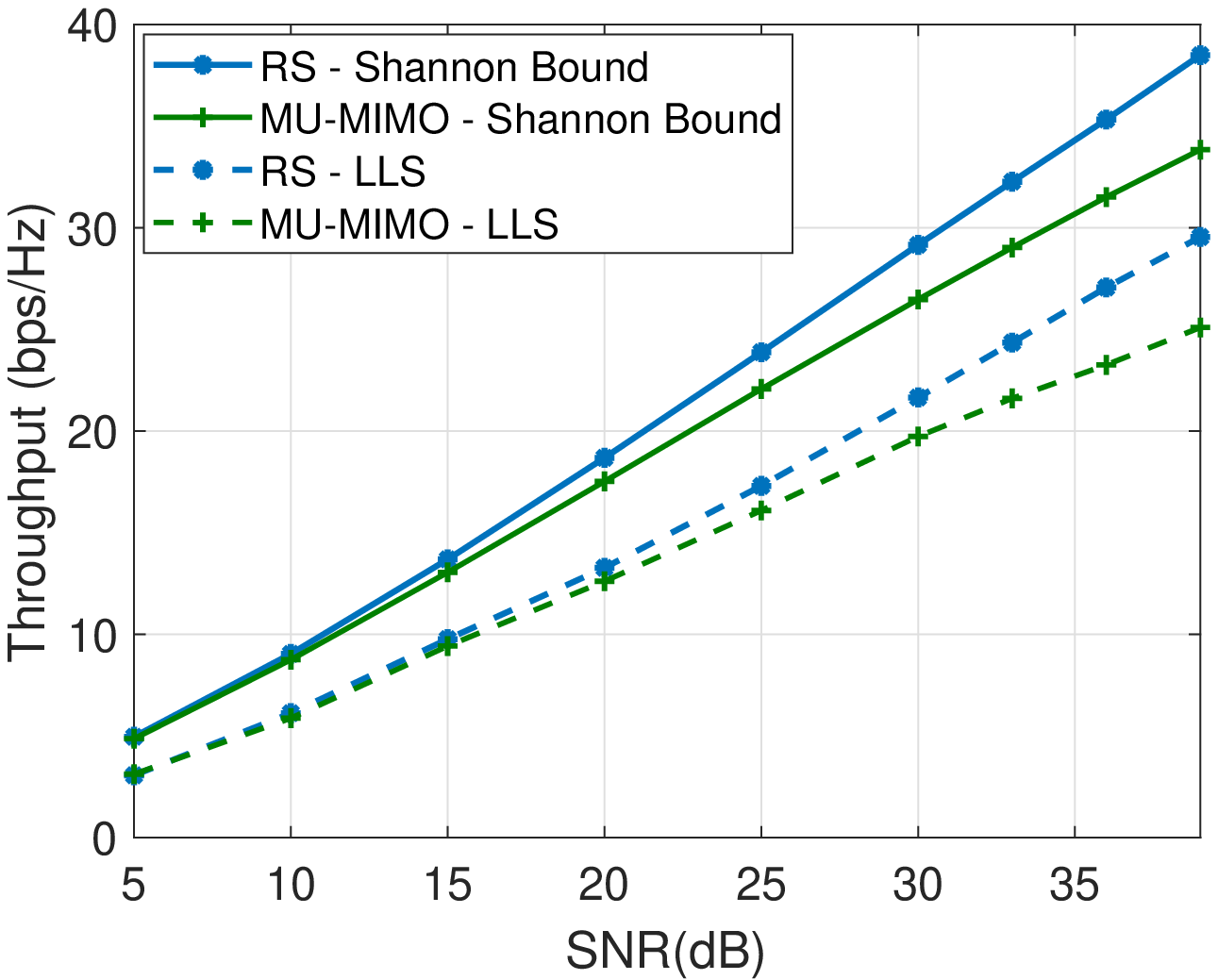}}%
\caption{Throughput versus SNR comparison of different strategies averaged over 100 random channel realizations.}
\label{fig:Throughput}
\end{figure}

In this section, we aim to verify the theoretical foundations in the paper under realistic and practical setups. We perform  Link-Level Simulations (LLS) to analyze the throughput performance of RS and compare it with those of MU-MIMO and MIMO NOMA. 
We employ the practical transceiver architecture described in Section~\ref{sec:transceiver} for RS, MU-MIMO and MIMO NOMA. Note that in the proposed architecture, the common signal is turned off to simulate MU-MIMO and one out of two private signals is turned off to simulate MIMO NOMA. We assume that the instantaneous channel is perfectly known at the transmitter for MCS selection.
Let $S^{(l)}$ denote the number of channel uses in the $l$-th Monte-Carlo realization and $D_{s,k}^{(l)}$ denote the number of successfully recovered information bits by user-$k$ in the common stream (excluding the common part of the message intended for the other user) and its private stream. Then, we calculate the throughput as  
\begin{align}
\mathrm{Throughput[bps/Hz]}=\frac{\sum_{l}(D_{s,1}^{(l)}+D_{s,2}^{(l)})}{\sum_{l}S^{(l)}}.
\end{align}
Fig.~\ref{fig:Throughput} shows the Shannon Bound (ESR obtained with Gaussian signalling and infinite block length) and throughput levels achieved by RS, MU-MIMO and MIMO NOMA in both underloaded and overloaded scenario, for $M=4, Q=2, Q_{c}=2$ and $\alpha=0.6$. The throughput performance is consistent with the ESR performance for all three schemes in both the underloaded and overloaded regime. 
\section{Conclusion}\label{Concl}
To conclude, we introduce a general framework for RSMA in MIMO BC with both perfect and imperfect CSIT, where RSMA is able to transmit an arbitrary number of common streams. We study the proposed framework in both finite and high SNR regimes. In the finite SNR regime, we propose the vectorization and WMMSE-based approach for precoder optimization to analyze the rate performance with perfect and imperfect CSIT. In the high SNR regime, we derive the sum-DoF achieved by RS, MU--MIMO and MIMO NOMA in a symmetric MIMO BC setup with imperfect CSIT. We demonstrate via the theoretical results that by transmitting multiple common streams, RSMA achieves a higher sum-DoF with imperfect CSIT compared to the conventional multiple access schemes. Numerical results show that RSMA achieves a higher rate performance than MU--MIMO and MIMO NOMA irrespective of the network load, user deployments or the CSIT quality. Moreover, numerical results validate the theoretical sum-DoF expressions derived for all the three transmission strategies and the sum-DoF gain of RS over MU--MIMO and MIMO MOMA. Moving beyond the assumptions of Gaussian signalling and infinite block lengths, we design the PHY-layer architecture of RS and analyze its performance in practical systems. Through LLS simulations, we demonstrate that by better managing the interference, RS achieves a significant throughput performance gain over MU-MIMO and MIMO NOMA in MIMO BC. Therefore, we conclude that RSMA is a powerful and promising physical-layer strategy for multi-antenna networks.
 
\appendices

\section{Achievable Sum-DoF}\label{AchSumDoF}
Since the precoders for each channel estimate are decoupled with each other, we consider precoding scheme for a given channel estimate $\widehat{\mathbf{H}}$ with precoders defined by $\{\mathbf{P}\}_{P_{t}}$. We begin with a general case by denoting the CSIT quality of the user-$k$ as ${\alpha_k}=\alpha,\; \forall\, k \in \mathcal{K}$. Here, ${\alpha}$ assumes a non-negative value and is in the range $[0,1]$. Also, replacing negative value of $\alpha$ with zero does not alter the sum-DoF results derived next. Therefore, one has $\mathbb{E}[{\lvert\mathbf{h}_{k,i}^{H}\mathbf{p}_k\rvert}^{2}]\sim P^{-\alpha}$,
with $\mathbf{h}_{k,i}$ as the $i^{th}$ column of channel ${\mathbf{H}_k}$ and $\mathbf{p}_k$ being the vector of unit norm in the null space of $\widehat{\mathbf{H}}_k$, i.e., ZF precoder. The entity ${\lvert\mathbf{h}_{k,i}\mathbf{p}_k\rvert}^{2}$ represents the residual interference power at the unintended receiver. The power exponent of the private stream vector of each user is taken as $\alpha=\min_{k\in \mathcal{K}}\{\alpha_k\}, \forall\,k\in\mathcal{K}$, as this power exponent   maximizes the sum-DoF \cite{chenxi2017bruno}. As aforementioned, here we consider a homogeneous network with equal number of receive antennas $Q$ at every user.
\subsection{RS and MU-MIMO}
We begin by first deriving the achievable sum-DoF of RS. We consider two possible scenarios, $M\leq Q$ and $M>Q$. In the former scenario, the sum-DoF is restricted to $M$ and can be achieved by switching to single user transmission. Therefore, we focus on the latter scenario where $M>Q$. This scenario encompasses both network load regimes, i.e., underloaded when $M=KQ$\footnote{In terms of the achievable sum-DoF, $M>KQ$ is equivalent to the $M=KQ$ scenario \cite{Elia2014}.} and overloaded when $M=LQ,\,1<L<K$. Consequently, we assume the number of users that will receive the private streams to be $J\leq K$ such that $M=JQ$. Whereas, the common streams will be multicast to all the $K$ users. Let us denote the set of users receiving the private streams to be $\mathcal{J}\subseteq \mathcal{K}$. Remaining $K-J$ users will form a set $\mathcal{J}^{*}$ such that $\mathcal{J}\cup \mathcal{J}^{*}=\mathcal{K}$. Therefore, we have $Q_{k}=Q,\,\forall k\in \mathcal{J}$ and $Q_{k}=0,\,\forall k\in \mathcal{J}^{*}$. We assume that the power is uniformly distributed among users for the private streams, i.e., $P^{\alpha}/J$ and consequently the power allocated to the common streams is $P-P^{\alpha}$, which  simply boils down to $P-P^{\alpha} \sim P$ in high SNR regimes. Hence, the transmission block is constructed as,
\begin{itemize}
\item A private stream vector denoted by $\mathbf{s}_k \in \mathbb{C}^{Q\times1}$ is transmitted to each user in the set $\mathcal{J}$ using a ZF precoder $\mathbf{P}_k \in \mathbb{C}^{M \times Q},\,\, \forall\, k\in \mathcal{J}$.
\item A common stream vector denoted by $\mathbf{s}_c \in \mathbb{C}^{Q_c \times 1}$ is multicast to all users using a precoder $\mathbf{P}_c \in \mathbb{C}^{M \times Q_c}$, where $\mathbf{P}_c$ constitutes the first $Q_c$ left singular vectors of $\widehat{\mathbf{H}}$.
\end{itemize}
As per the system model, the transmitted signal from the BS and the received signal at user-$k$ writes as
\begin{equation}{\label{eq:DoF11}}
\mathbf{x}=\underbrace{\mathbf{P}_{c}\mathbf{s}_c}_{P}+\sum_{k\in\mathcal{J}}\underbrace{\mathbf{P}_{k}\mathbf{s}_{k}}_{P^{\alpha}/J},
\end{equation}
\begin{equation}{\label{eq:DoF12}}
\mathbf{y}_{k,k\in\mathcal{J}}=\underbrace{\mathbf{H}_k^{H}\mathbf{P}_{c}\mathbf{s}_{c}}_{P}+\underbrace{\mathbf{H}_k^{H}\mathbf{P}_{k}\mathbf{s}_k}_{P^{\alpha}}+\sum_{j\in\mathcal{J},j\neq k}\underbrace{\mathbf{H}_k^{H}\mathbf{P}_{j}\mathbf{s}_{j}}_{P^{\alpha-\alpha_k}}+\underbrace{\mathbf{n}_k}_{P^{0}}.
\end{equation}
\begin{equation}{\label{eq:DoF13}}
\mathbf{y}_{k,k\in\mathcal{J}^{*}}=\underbrace{\mathbf{H}_k^{H}\mathbf{P}_{c}\mathbf{s}_{c}}_{P}+\sum_{j\in\mathcal{J}}\underbrace{\mathbf{H}_k^{H}\mathbf{P}_{j}\mathbf{s}_{j}}_{P^{\alpha}}+\underbrace{\mathbf{n}_k}_{P^{0}}.
\end{equation}
\subsubsection{User-$k$ in $\mathcal{J}$} First, we write the common and private rates achieved by user-$k$,

\begin{equation}{\label{eq:DoF4}}
\widehat{R}_{c,k}= \log_{2}\det(\mathbf{Q}_{c}+\mathbf{Q}_{k}+\mathbf{Q}_{\eta_k})-\log_{2}\det(\mathbf{Q}_{k}+\mathbf{Q}_{\eta_k}),
\end{equation}
\begin{equation}{\label{eq:DoF5}}
\widehat{R}_{p,k}= \log_{2}\det(\mathbf{Q}_{k}+\mathbf{Q}_{\eta_k})-\log_{2}\det(\mathbf{Q}_{\eta_k}),
\end{equation}
where $\mathbf{Q}_{c}=\mathbf{{H}}_{k}^{H}\mathbf{P}_c\mathbf{P}_{c}^{H}\mathbf{{H}}_{k}$, $\mathbf{Q}_{k}=\mathbf{{H}}_{k}^{H}\mathbf{P}_k\mathbf{P}_{k}^{H}\mathbf{{H}}_{k}$ and $\mathbf{Q}_{\eta_k}=\sum_{j\in\mathcal{J},j\neq k}\mathbf{{H}}_{k}^{H}\mathbf{P}_j\mathbf{P}_{j}^{H}\mathbf{{H}}_{k}+\sigma_{n}^{2}\mathbf{I}_{Q}$ are respective covariance matrices of $\mathbf{{H}}_{k}^{H}\mathbf{P}_c\mathbf{s}_{c}$, $\mathbf{{H}}_{k}^{H}\mathbf{P}_k\mathbf{s}_{k}$ and $\sum_{j\in\mathcal{J},j\neq k}\mathbf{{H}}_{k}^{H}\mathbf{P}_j\mathbf{s}_{j}+\mathbf{n}_{k}$. 
Furthermore, we consider the eigenvalue decomposition of $\mathbf{Q}_{c}$, $\mathbf{Q}_{k}$ and $\mathbf{Q}_{\eta_k}$ as $\mathbf{W}_c\mathbf{D}_c\mathbf{W}_{c}^{H}$, $\mathbf{W}_k\mathbf{D}_k\mathbf{W}_{k}^{H}$ and  $\mathbf{W}_{\eta_k}\mathbf{D}_{\eta_k}\mathbf{W}_{\eta_k}^{H}$, respectively, with $\mathbf{D}_c \sim diag(P\mathbf{I}_{Q_c},\mathbf{0}_{Q-Q_{c}})$, $\mathbf{D}_{k} \sim diag(P^{\alpha}\mathbf{I}_{Q})$ and  $\mathbf{D}_{\eta_k} \sim diag(P^{(\alpha-\alpha_k)^{+}}\mathbf{I}_{Q})$. Here, $(x)^{+}\triangleq \max(x,0)$. Then it follows,
\begin{equation}{\label{eq:DoF8}}
\begin{split}
\log_{2}\det(\mathbf{Q}_{c}+\mathbf{Q}_{k}+\mathbf{Q}_{\eta_k})&=\big(Q_c+(Q-Q_c)\alpha\big)
\log_{2}(P_t)+\mathcal{O}\,(\log_2(P_t)),
\end{split}
\end{equation}
\begin{equation}{\label{eq:DoF9}}
\begin{split}
\log_{2}\det(\mathbf{Q}_k+\mathbf{Q}_{\eta_k})&=(Q\alpha)
\log_{2}(P_t)+\mathcal{O}\,(\log_{2}(P_t)),
\end{split}
\end{equation}
\begin{equation}{\label{eq:DoF10}}
\log_{2}\det(\mathbf{Q}_{\eta_k})=\mathcal{O}\,(\log_{2}(P_t)).
\end{equation}
\subsubsection{User-$k$ in $\mathcal{J}^{*}$} For user-$k$ in the the set $\mathcal{J}^{*}$, we only have the common stream vector and the common rate is written as 
\begin{equation}{\label{eq:DoFO1}}
\widehat{R}_{c,k}= \log_{2}\det(\mathbf{Q}_{c}+\mathbf{Q}_{\eta_k})-\log_{2}\det(\mathbf{Q}_{\eta_k}),
\end{equation}
where $\mathbf{Q}_{\eta_k}=\sum_{j\in\mathcal{J}}\mathbf{{H}}_{k}^{H}\mathbf{P}_j\mathbf{P}_{j}^{H}\mathbf{{H}}_{k}+\sigma_{n}^{2}\mathbf{I}_{Q}$ with the eigenvalue decomposition as $\mathbf{D}_{\eta_k} \sim \\ diag(P^{\alpha}\mathbf{I}_{Q})$. It follows 
\begin{equation}{\label{eq:DoFO2}}
\begin{split}
\log_{2}\det(\mathbf{Q}_{c}+\mathbf{Q}_{\eta_k})&=\big(Q_c+(Q-Q_c)\alpha\big)
\log_{2}(P_t)+\mathcal{O}\,(\log_2(P_t)),
\end{split}
\end{equation}
\begin{equation}{\label{eq:DoFO4}}
\log_{2}\det(\mathbf{Q}_{\eta_k})=(Q\alpha)\,\log_{2}(P_t).
\end{equation}
The term $\mathcal{O}(\log_2 (P_t))$ dies with $P_t \rightarrow \infty$ because of the negative exponent. We proceed to obtaining the common and private DoF for user-$k$ by using the expression in (\ref{eq:b4}). From (\ref{eq:DoF4}), (\ref{eq:DoFO1}), (\ref{eq:DoF8})--(\ref{eq:DoF9}) and (\ref{eq:DoFO2})--(\ref{eq:DoFO4}), the DoF for the common stream vector is $d_{c,k}=Q_{c}(1-\alpha)$ and is shared by all users in $\mathcal{K}$. Similarly, from (\ref{eq:DoF5}) and (\ref{eq:DoF9})--(\ref{eq:DoF10}), the DoF for the private stream vector at each user in $\mathcal{J}$ is $d_{p,k}=Q\alpha$ and $d_{p,k}=0$ for users in $\mathcal{J}^{*}$. Considering the total rate achieved by user-$k$, $\forall k\in\mathcal{K}$, it can be written that  $\widehat{R}_{c}+\widehat{R}_{p,k}\leq \widehat{R}_{c,k}+\widehat{R}_{p,k}$. Hence, the sum-DoF achieved by RS is $d_{s}^{\textrm{RS}}\leq d_{c,k}+\sum_{k=1}^{K}d_{p,k}$ and is expressed in (\ref{eq:4}).
\par Switching off the common stream and using equations (\ref{eq:DoF5})-(\ref{eq:DoF10}), the expression for the sum-DoF achieved by the conventional MU--MIMO scheme is obtained and is expressed in (\ref{eq:5}).
\subsection{MIMO NOMA} 
We now consider multi-antenna MIMO NOMA. Without loss of generality we assume the decoding order\footnote{The sum-DoF analysis is independent of the basis of ordering and will hold for any ordering of users.} $1\rightarrow K$ such that user-$1$ performs $K-1$ layers of SIC to fully decode the messages (and therefore remove interference) from the other $K-1$ users. Similarly, the next user, i.e., user-$2$ performs $K-2$ layers of SIC to fully decode messages from other $K-2$ users and so on. Thus, user-$K$ decodes its own message vector treating messages of the rest $K-1$ users as noise and the message vector of user-$1$ will be decoded by user-$1$ after it decodes the messages of all the other $K-1$ users and removes them. Following the NOMA principle, the transmit signal vector $\mathbf{x}$ is generated such that the messages intended for user-$k$ are encoded using a shared codebook such that user-$k$,$\; k \in \mathcal{K}$ is able to decode the message of user-$j,\; j\in \mathcal{K}$ for $j>k$. After encoding the messages, linear precoders $\mathbf{P}_k \in \mathbb{C}^{M\times Q_{k}},\; \forall k \in \mathcal{K}$ can be used to construct the transmit signal vector denoted by 
\begin{equation}{\label{eq:NOMA_x}}
\mathbf{x}=\sum_{k\in\mathcal{K}}\underbrace{\mathbf{P}_{k}\mathbf{s}_{k}}_{P/K},
\end{equation}
and the received signal at user-$k$ is given by $\mathbf{y}_k=\mathbf{H}_{k}\mathbf{x}+\mathbf{n}_k$. Using SIC, user-$k$ decodes the message vector of user-$j,\; j\geq k$ while treating the interference from users $\{i\;\mid i < j,\; i \in \mathcal{K}\}$ as noise. Under the assumption of Gaussian signalling and perfect SIC, the rate at user-$k$, to decode the message vector of user-$j$ is given by
\begin{equation}
\widehat{R}_{k,j}=\log_{2}\det\bigg(\mathbf{I}_{Q}+{\mathbf{Q}_{k,j}}\big(\mathbf{I}_{Q}+\mathbf{Q}_{k,j}^{(in)}\big)^{-1}\bigg),
\end{equation}
where $\mathbf{Q}_{k,j}=\mathbf{{H}}_{k}^{H}\mathbf{P}_j\mathbf{P}_{j}^{H}\mathbf{{H}}_{k}$ and $\mathbf{Q}_{k,j}^{(in)}=\sum_{i<j,i \in \mathcal{K}}\mathbf{{H}}_{k}^{H}\mathbf{P}_i\mathbf{P}_{i}^{H}\mathbf{{H}}_{k}$. In order to ensure decodability, message vector of user-$j$ is expected to be decoded at user-$k\;\mid j\geq k,\; k \in \mathcal{K}$ and thus the rate of user-$j$ should not exceed $\widehat{R}_{j}=\min_{k \leq j,\; k \in \mathcal{K}}\,\widehat{R}_{k,j}$.
Therefore, the achievable rates of $K$-users is given by $\widehat{R}_{1}=\widehat{R}_{1,1}$, $\widehat{R}_{2}= \min(\widehat{R}_{1,2},\widehat{R}_{2,2})$,
$\widehat{R}_{3}= \min(\widehat{R}_{1,3},\widehat{R}_{2,3},\widehat{R}_{3,3})$,\ldots, $\widehat{R}_{K}= \min(\widehat{R}_{1,K},\widehat{R}_{2,K},\ldots,\widehat{R}_{K,K})$. Since the message vector of each user is required to be decoded by user-$1$, the SR $\widehat{R}_{\textrm{NOMA}}$ of the $K$-user MIMO NOMA can then be upper bounded as
\begin{equation}
\begin{split}
\widehat{R}_{\textrm{NOMA}} &\leq \sum_{k=1}^{K} \widehat{R}_{1,k}=\log_{2}\det(\mathbf{I}_{Q}+\sum_{k=1}^{K}\mathbf{Q}_{1,k}).
\end{split}
\end{equation}
The SR bound achieved with this $K$-user MIMO NOMA strategy is further upper bounded as
 \begin{align}
\widehat{R}_{\textrm{NOMA}}&\leq \log_2\det\left(\mathbf{I}_Q+\mathbf{H}_1^H \mathbf{Q}_1^{\star} \mathbf{H}_1\right)\nonumber\stackrel{P_t\nearrow}{=} \min(M,Q)\log_2\left(P_{t}\right) + \mathcal{O}(1),\label{eq_MIMO_OMA}
\end{align}
where $\mathbf{Q}_1^{\star}$ refers to the optimal covariance matrix for user-1 in a single-user (OMA) setup with $tr(\mathbf{P}_1\mathbf{P}_{1}^{H})= P_{t}$, i.e., obtained by transmitting along the dominant eigenvector of $\mathbf{H}_1\mathbf{H}_1^H$ and allocating power $P_t$ according to the water-filling solution. The sum-DoF obtained for MIMO NOMA is described in equation (\ref{eq:nomadof}).
\ifCLASSOPTIONcaptionsoff
  \newpage
\fi
\bibliographystyle{IEEEtran}
\bibliography{refer}
\end{document}